\documentclass[%
reprint,
superscriptaddress,
nofootinbib,
 amsmath,amssymb,
 aps,
 prd,
floatfix,
]{revtex4-1}
\pdfoutput=1

\usepackage{graphicx}
\usepackage{epstopdf}
\usepackage{dcolumn}
\usepackage[hidelinks]{hyperref}
\usepackage{cleveref}
\usepackage{physics} 
\usepackage{mathtools}
\usepackage[normalem]{ulem}
\usepackage{color}
\newcommand{\N}{\mathbb{N}}
\DeclareMathOperator{\Var}{Var}
\newcommand{\change}[1]{#1}
\crefname{section}{Sec.}{Secs.}
\crefname{figure}{Fig.}{Figs.}
\crefname{equation}{Eq.}{Eqs.}
\crefname{enumi}{assumption}{assumptions}
\newlength{\figwidth}
\begin{document}

\title{Can the fluctuations of the quantum vacuum solve the cosmological constant problem?}

\author{Samuel S.\ Cree}
 \email{samuel.cree@uq.net.au}
\author{Tamara M.\ Davis}
\author{Timothy C.\ Ralph}
\affiliation{The University of Queensland, School of Mathematics and Physics, QLD 4072, Australia}
\author{Qingdi Wang}
\author{Zhen Zhu}
\author{William G.\ Unruh}
\affiliation{Department of Physics and Astronomy, The University of British Columbia, Vancouver V6T 1Z1, Canada}
\date{\today}

\begin{abstract}
	The cosmological constant problem arises because the magnitude of vacuum energy density predicted by quantum mechanics is about $120$ orders of magnitude larger than the value implied by cosmological observations of accelerating cosmic expansion.
Recently some of the current authors proposed that the stochastic nature of the quantum vacuum can resolve this tension \cite{wzu}.  
By treating the fluctuations in the vacuum seriously and allowing fluctuations up to some high-energy cutoff at which Quantum Field Theory is believed to break down, a parametric resonance effect arises that leads to a slow expansion and acceleration. 
In this work we thoroughly examine the implications of this proposal by investigating the resulting dynamics.  
Firstly, we improve upon numerical calculations in the original work and show that convergence issues had overshadowed some important effects.
Correct calculations reverse some of the conclusions in \cite{wzu}, however the premise that parametric resonance can explain a very slowly accelerating expansion appears to remain sound.  
After improving the resolution and efficiency of the numerical tests, we explore a wider range of cutoff energies, and examine the effects of multiple particle fields.  
We introduce a simple model using the Mathieu equation (a prototypical example of parametric resonance), and find that it closely matches numerical results in regimes where its assumptions are valid.
Using this model, we extrapolate to find that in a Universe with $28$ bosonic fields and a high-energy cutoff $40$ times higher than the Planck energy, the acceleration would be comparable to what is observed.
\end{abstract}


\maketitle


\section{Introduction}

One of the greatest challenges in modern physics is to reconcile general relativity and quantum physics into a unified theory.
Perhaps the most dramatic clash between the two theories lies in the cosmological constant problem \cite{weinberg,carroll,martin,dolgov,susskind,summary}. 
Naive predictions of vacuum energy from quantum physics predict a magnitude so high that the expansion of the Universe should have accelerated so quickly that no structure could have formed.
The predicted rate of acceleration resulting from vacuum energy is famously $120$ orders of magnitude larger than what is observed.

In a 2017 paper \cite{wzu} some of the current authors proposed a solution to the cosmological constant problem.
They proposed that rather than use the expectation value of the quantum energy density in the Einstein equations, which would lead to the overwhelmingly large prediction for cosmic acceleration, one should instead treat the vacuum as an inhomogeneous stochastic field.
Accounting for the fluctuations in the density of the vacuum energy---which are on the order of the magnitude of the vacuum energy itself---can potentially explain a slow expansion. 

Here we investigate that proposal with improved computational methods. 
Our faster computational methods allow us to make a more thorough investigation of the behavior of the expansion of the Universe in the presence of a stochastic vacuum field by extending the model to a greater number of particle fields. 
We find that the original calculations had not sufficiently converged, and a more thorough calculation reverses some of the trends seen in the original paper.
When these are remedied, the original proposal no longer explains the results when there are just two massless scalar particle fields in the Universe.
However, given that the standard model has dozens of particles, and $28$ bosonic field components
, a realistic model should contain many fields.
Our faster computational methods allow us to extend the model to a greater number of particle fields. 
With at least three fields, the exponentially small acceleration predicted by the original proposal is observed, and the magnitude of the acceleration gets smaller as more fields are added and the cutoff increased---meaning that with a sufficient number of fields at a sufficiently high cutoff, the predicted acceleration would match observation.
 
The paper is structured as follows.
In \Cref{sect:ccprob}, we summarize the aspects of the cosmological constant problem that are relevant to this work.
In \Cref{sect:stochastic}, we summarize the model of cosmological dynamics in the presence of a stochastic inhomogeneous vacuum that was introduced in \cite{wzu}, and how it attempts to resolve the problem.
We also mention a caveat to the application of the adiabatic theorem in \cite{wzu}, which implies that the resultant analytical description is only valid with three or more scalar fields present.
In \Cref{sect:sims}, we describe our numerical methods, which are similar to those used in \cite{wzu}, before testing the convergence of our new results across all relevant parameters to demonstrate that they are robust to all limits.
In \Cref{sect:results}, we provide corrections to numerical findings of the original paper, before using our improved methods to test greater numbers of particle fields, and a larger range of choices of cutoff frequency for vacuum oscillations.
Finally, we conclude with the physical significance of the new results in \Cref{sect:conclusion}.
Throughout, we use $\hbar=G=c=1$, and a metric signature of $(-,+,+,+)$.

\section{The Cosmological Constant Problem}\label{sect:ccprob}
In the Einstein equations of general relativity, a term representing the curvature of spacetime ($R_{\mu\nu}$) is related to a term describing the energy-momentum of matter ($T_{\mu\nu}$), as well as the cosmological constant $(\lambda)$ and metric tensor $(g_{\mu\nu})$ as follows:
\begin{align}
	R_{\mu\nu}-\frac{1}{2}R^{\sigma}_\sigma g_{\mu\nu} + \lambda g_{\mu\nu} &= {8\pi}T_{\mu\nu} \label{eq:Einstein}.
\end{align}
Each element of the curvature tensor and metric tensor are just classical fields, but the elements of the energy-momentum tensor must be quantum operators in order to account for known quantum effects of matter.
A currently undiscovered theory of quantum gravity would presumably elevate the left-hand side to become quantum operators. 
In the meantime it is common to treat both sides as classical (known as ``semiclassical'' gravity).
The most common way of doing this is to replace $T_{\mu\nu}$ with $\langle\hat{T}_{\mu\nu}\rangle$ (the Moller-Rosenfeld approach) \cite{rosenfeld,qftcs,qftcs2}.
But this approach fails in a number of ways: it allows faster-than-light communication \cite{evproblems2}, it leads to a nonlinear Hamiltonian which contradicts the Born rule \cite{evproblems}, and most infamously, it predicts an overwhelming large accelerating expansion of the Universe.

Here we outline the traditional approach to the cosmological constant problem, see \cite{weinberg,carroll,martin,straumann,positivelambda,padmanabhan}.
The usual argument states that the vacuum state $\ket{0}$ should be locally Lorentz invariant so that observers agree on the vacuum state.
This means that the expectation value of the energy-momentum tensor on the vacuum, $\bra{0} \hat{T}_{\mu\nu}\ket{0}$, must be a scalar multiple of the metric tensor $g_{\mu\nu}$ (which is the only Lorentz invariant rank $(0,2)$ tensor).
Because the $\hat T_{00}$ component is an energy density, we label $\bra{0} \hat T_{00}\ket{0} =\rho_{\textrm{vac}}$, so that the vacuum contribution to the right-hand side of \Cref{eq:Einstein} can be written
\begin{align}
	\bra{0} \hat T_{\mu\nu}\ket{0} = -{\rho_{\textrm{vac}}} g_{\mu\nu}.
	\label{eq:VacuumTensor}
\end{align}
Subtracting this from the right-hand side of \Cref{eq:Einstein} and grouping it with the cosmological constant term replaces $\lambda$ with an ``effective'' cosmological constant:
\begin{align}
	\lambda_{\textrm{eff}} &= \lambda + 8\pi {\rho_{\textrm{vac}}}.
	\label{eq:lambda}
\end{align}
\change{
	The meaning of \Cref{eq:VacuumTensor} is revealed by noticing that in flat spacetime (where $g_{\mu\nu} = \mathrm{diag}(-1,1,1,1)$), it implies $\rho_{\textrm{vac}}=-P_{\textrm{vac}}$, where $P_{\textrm{vac}}=\bra{0}\hat{T}_{ii} \ket{0}$ (for any $i\in \left\{ 1,2,3 \right\}$) is the pressure.
	Importantly, this implies that if the energy density is positive (as is usually assumed) then the pressure must be negative, a conclusion which extends to any metric $g_{\mu\nu}$ with a (1,3) signature.
	Whereas gravity is attractive \footnote{\change{By ``attractive'', we mean that the strong energy condition is satisfied (i.e. $\Omega^2>0$, where $\Omega^2$ is defined by \Cref{eq:WzuOmega}), and that gravity has a tendency to pull things closer together. 
	A more specific example is this: given a small ball of freely falling test particles initially at rest with respect to each other, gravity is attractive if the second derivative of the volume of the ball is negative, i.e. the ball tends to shrink (more details about this picture are described in \cite{attractive}).}} for positive energy and pressure, in conditions with strong negative pressure it becomes repulsive, which leads to accelerating expansion.

	The usual method of quantifying the accelerating expansion comes from describing the Universe with the Friedmann-Robertson-Walker metric:
	\begin{align}
		\mathrm{d} s^2 = -\mathrm{d} t^2 + a(t)^2 \left( \mathrm{d} x^2+ \mathrm{d} y^2+ \mathrm{d} z^2 \right),
		\label{eq:HomMetric}
	\end{align}
in which the scale factor $a$ represents the overall scale of the Universe.
This is the standard metric used in cosmology, and is known to accurately represent the Universe on large scales.}
Then the relative acceleration of the expansion of the Universe ($\frac{\ddot{a}}{a}$, where a dot denotes a time derivative), is found to be directly proportional to the effective cosmological constant, and is measured to be about $10^{-122}$ in Planck units.

Now, we determine ${\rho_{\textrm{vac}}}$ in the simplified case of a single massless spin-$0$ particle field.
For each 4-momentum $k=\left( \omega_k, \vb k \right)$, the field acts like a simple harmonic oscillator.
The $n$th state, with energy $\left( n+\frac{1}{2} \right)\omega_k$ (recalling $\hbar=1$), contains $n$ particles with momentum $\vb k$ and energy $\omega_k$, and the ground state (with no particles) has energy $\frac{1}{2}\omega_k$.
Combining the ground state energy of each mode (i.e.\ the harmonic oscillator that corresponds to each 4-momentum) yields an infinite value for the vacuum energy density.
By restricting to modes with particle energy below a certain cutoff energy $\omega_k \le \Lambda$ (not to be confused with $\lambda$, the cosmological constant), a finite, regularized result for the energy density can be obtained. 
The result is proportional to $\Lambda^4$, because the number of allowed modes scales with $\Lambda^3$, and the average energy of the allowed modes scales linearly with $\Lambda$.
Any other fields will contribute similarly, so that if there are $n_{\mathrm{f}}$ scalar fields, the density scales with $n_{\mathrm{f}} \Lambda^4$.
Typically, the cutoff is taken to be near $\Lambda=1$ in Planck units (i.e.\ the Planck energy), so the vacuum energy gives a contribution to the cosmological constant on the order of at least unity according to \Cref{eq:lambda}.
Thus we see the extreme fine-tuning problem: the original cosmological constant $\lambda$ must cancel this large vacuum energy density ${\rho_{\textrm{vac}}}\sim 1$ to a precision of $1$ in $10^{120}$---but not completely---to result in the observed value $\lambda_{\textrm{eff}} \sim 10^{-120}$.

\section{Cosmological Dynamics under Semiclassical Stochastic Gravity}\label{sect:stochastic}
The energy density of the vacuum state fluctuates wildly, with variations comparable to its magnitude.
Thus, rather than ignoring these fluctuations by treating the vacuum energy density as constant, some of the current authors \cite{wzu} proposed treating it as an inhomogeneous stochastic field to better approximate a full quantum description. 
\change{

	Three key changes are made to the traditional approach outlined above.
	First, in order to allow spatial variations and inhomogeneity, \Cref{eq:HomMetric} is replaced with the following metric:
	\begin{align}
		\mathrm{d} s^2 = -\mathrm{d} t^2 + a(t,\vb x)^2 \left( \mathrm{d} x^2+ \mathrm{d} y^2+ \mathrm{d} z^2 \right),
		\label{eq:InhomMetric}
	\end{align}
	i.e.\ the scale factor $a(t,\vb x)$ is now inhomogeneous, representing the relative ``size'' of spacetime at each point.
	It was noted in \cite{wzu} that solving one of the Einstein equations for $a(t,\vb x)$ in this metric   dooes not necessarily mean you can solve the rest of them simultaneously.
	One degree of freedom on the left-hand side of \Cref{eq:Einstein} will not capture the complexities on the right-hand side.
	One could use a more general inhomogeneous metric but the equations become far more difficult to solve.

	Secondly, the right-hand side of \Cref{eq:Einstein} is treated not as an expectation value but as a stochastic inhomogeneous field that acts as a source for these inhomogeneities, in a manner that will be clarified shortly.
	Other semiclassical stochastic gravity approaches have been considered before \cite{hu,stochastic}, but quite differently to what is presented here.
	The main difference between our work and theirs is that we couple both the huge expectation value and the fluctuations of the zero point energy to gravity without trying regularization methods to make them small; they consider the fluctuations in quantum fields but they disregard the huge expectation value and try regularization to make the fluctuations small. 

	Finally, we do not assume Lorentz invariance, so that \Cref{eq:VacuumTensor} no longer holds. 
	Instead we assume an explicit cutoff in frequency and, as above, we assume an explicitly non-Lorentz invariant form of the metric.
	Both the energy density and the pressure are large and positive, and the the matter gravitates attractively (as defined above).
	The physical justification of this last assumption will be discussed further in \Cref{sect:conclusion}, but we will summarize this discussion here:
	\begin{itemize}
		\item The high-energy cutoff $\Lambda$ used to determine $\rho_{\textrm{vac}}$ in \Cref{sect:ccprob} inherently violates local Lorentz invariance already, so using this in combination with \Cref{eq:VacuumTensor} is inconsistent \cite{martin,covariant1,covariant2,covariant3}.
	\item Many proposals for quantum gravity suppose some discrete spacetime structure arises at a small invariant length scale. Such a length scale must violate Lorentz invariance \cite{lengthscale,foam,foam1,foam2}.
	\end{itemize}

	Because we do not require \Cref{eq:VacuumTensor} to hold, our model of the vacuum no longer has negative pressure when energy density is positive, so gravity can be attractive everywhere.
	It has been shown in \cite{wzu} that even with gravity being purely attractive, our model still predicts apparent ``repulsive'' effects (a slow exponential expansion) on large scales.
	This arises from attractive gravity due to the parametric resonance effect---a harmonic oscillator is always ``attracted'' towards its equilibrium point but its swing amplitude (which represents the size of space) grows exponentially.
	However on intermediate scales (much larger than the cutoff scale but smaller than the cosmological scale) fields act as though they are on a Lorentz invariant spacetime, as shown in \cite{wzu}.

	Now, by applying these assumptions we can use the Einstein equations (which we assume still hold) to determine the dynamics of the evolution of the Universe.
	An alternative but equivalent expression to \Cref{eq:Einstein} is the following:
\begin{align}
	R_{\mu\nu} - \lambda g_{\mu\nu} &= {8\pi} \left( T_{\mu\nu} - \frac{1}{2}T_{\sigma}^\sigma g_{\mu\nu} \right) \label{eq:EinsteinAlternative}.
\end{align}
The key dynamical equation that we use in this work arises from the $\mu=0,\ \nu=0$ equation (the ``temporal'' equation, because it contains only time derivatives of $a(t,\vb x)$), with $\lambda=0$.
It takes the following form:
}
\begin{align}
	\ddot a(t,\vb x) + \Omega^2(t,\vb x) a(t,\vb x) = 0.
\label{eq:WzuDynamics}
\end{align}
	We can recognize \Cref{eq:WzuDynamics} as a harmonic oscillator equation for each $\vb x$, with $\Omega$ playing the role of a frequency (not to be confused with the usual use of $\Omega$ in cosmology to mean energy density).
The square of the frequency of those oscillations is proportional to a linear combination of components of the energy-momentum tensor, which we treat as time and position dependent stochastic fields:
\change{
\begin{align}
	\Omega^2(t,\vb x) = \frac{4\pi}{3} \left( \rho(t,\vb x) + \sum_{i=1}^{3} P_i(t,\vb x) \right),
\label{eq:WzuOmega}
\end{align}
where $\rho(t,\vb x)=T_{00}(t,\vb x)$ and $P_i(t,\vb x) = a(t,\vb x)^{-2} T_{ii}(t,\vb x)$ are both stochastic fields.

These stochastic fields are chosen according to the operators $\hat{T}_{\mu\nu}$ and the vacuum state $\ket{0}$.
The simple model that we will use arises from choosing to use a number $n_{\textrm{f}}$ of massless scalar fields (which was just one for most of \cite{wzu}, but here we will extend this to a greater number of fields).}
\change{For a massless scalar field $\phi$, it happens that $\Omega^2=\frac{8\pi}{3}\dot\phi^2$ is independent of $a$ and strictly positive.} 
This need not always be true for massive, or fermionic fields, because they add negative terms to the expression for $\Omega^2$.
$\Omega^2$ being strictly positive means that ${\Omega} \equiv {\sqrt{\Omega^2}}$ is well defined, and that \Cref{eq:WzuDynamics} will always act like a harmonic oscillator, rather than yield an explicitly exponential solution (like, for example, $\ddot{a}=a$).

\change{As in \cite{wzu}, we continue to use the high-energy cutoff regularization approach that was introduced in \Cref{sect:ccprob}. 
	Although this method violates local Lorentz invariance, there are other regularization methods that do not, and their effects on this new proposal have been discussed in \cite{comment,reply,fermionic} (namely that they do not always lead to a positive definite expression for $\Omega^2$).
We continue to use the high-energy cutoff method here because we do not believe that the Lorentz-invariant methods are physical representations of the huge vacuum energy implied by zero-point fluctuations, and because the high-energy cutoff has physical meaning as per the effective field theory interpretation discussed in \Cref{sect:conclusion}.}

Once $T_{\mu\nu}$ is defined according to the choice of fields and regularization method, we determine the stochastic properties of $\Omega$ (expecation value, variance, power spectrum, etc.)~by considering the $T_{\mu\nu}$ components in \Cref{eq:WzuOmega} as classical stochastic fields, whose statistical properties are described by vacuum expectations (e.g.\ variance $\bra{0}\Omega^4\ket{0}-\bra{0}\Omega^2\ket{0}^2$).
Because we are only considering the vacuum, and no excitations, the cosmological scenario being described is a simplified model consisting only of vacuum energy.
If $T_{\mu\nu}$ contained contributions from all the fields in our Universe, this would be approximately equivalent to studying our own Universe in the current, dark-energy-dominated epoch.

The vacuum state is not an eigenstate of the local energy density and pressure operators in \Cref{eq:WzuOmega}, so measurements of these variables will fluctuate with a predictable spectrum.
By modeling these fluctuations stochastically, $\Omega$ becomes a quasiperiodic function in space and time---meaning that its statistical properties are constant, but there is no fixed period $T$ for which $\Omega(t,\vb  x) = \Omega(t+T,\vb  x)$ or $\vb  X$ for which $\Omega(t,\vb  x) = \Omega(t,\vb  x+\vb  X)$, as would be the case for a strictly periodic function.

Solutions to harmonic oscillator equations with time-dependent frequency, like this one, can exhibit long-term growth or decay, a phenomenon known as parametric resonance \cite{landau,parresstability}.
A common example of parametric resonance occurs on a swing, when one straightens and bends one's legs to increase the amplitude.
Because of the linearity and symmetry of \Cref{eq:WzuDynamics}, it turns out that decaying solutions will be suppressed unless the initial conditions are fine-tuned, so that the long-term solution will either grow exponentially or remain steady.
This means that the general solution can be written as,
\begin{align}
	\change{a(t, \vb x)\approx e^{Ht}P(t, \vb x),}
	\label{eq:WzuSolution}
\end{align}
where $H\ge 0$ is a constant and $P$ is a quasiperiodic function, by which we mean that all its statistical properites are time-independent, and it has time average $ \overline{P} = 0$.
Note that we use, for example, $\overline{P}$ to denote the time average of a variable, reserving $\ev{P}$ to denote the expectation of $P$ as a quantum operator.
Note that $\frac{\dot{P}}{P} = \frac{\dd \log|P|}{\dd t}$, so since $\overline{\log|P|}=$const., $\overline{( \frac{\dot{P}}{P} )}=0$.
This means that taking a time average of $\frac{\dot{a}}{a} = H  + \frac{\dot{P}}{P}$ gives us $H = \overline{\left( \frac{\dot{a}}{a} \right)} $.
This leads to a natural interpretation of $H$ as the Hubble parameter, which is defined in cosmology as $\frac{\dot{a}}{a}$.
If $H$ is zero, then there is no parametric resonance, because $a=P$ and $P$ has no long-term growth or decay.
Otherwise, it will result in an exponentially increasing scale factor, resulting in observed distances scaling with $L(t) = L(0) e^{Ht}$, and macroscopic acceleration obeying $ {\frac{\ddot{L}(t)}{L(t)}}=H^2$.

Thus, the key goal is to determine $H^2$; if $H^2 \sim 1$, the model has done nothing to remedy the problem of the traditional approach, as it still predicts an acceleration $120$ orders of magnitude too large.
If $H^2\sim 10^{-120}$, then this would indicate that the model predicts an appropriate order of magnitude for the acceleration, and has potential to resolve the cosmological constant problem.

\subsection{Timescales of oscillation}
Parametric resonance is usually strongest (i.e.\ growth or decay is most rapid) when the timescale of frequency oscillation and amplitude oscillation are similar---e.g.\ when one bends one's legs with a frequency close to the frequency of the swing itself.
It is, therefore, important to assess the conditions under which the variations in $\Omega$ are of a similar frequency to those of $a$, since that is when accelerating expansion will be strongest.
This will also provide us with expectations of the limiting behavior when the oscillations in $\Omega$ are much slower or faster than those of $a$.
Because \Cref{eq:WzuDynamics} contains no spatial derivatives, we will omit the label $\vb  x$ and just consider a fixed point in space from now on.

Although neither $\Omega(t)$ nor $a(t)$ are strictly periodic, their variations still occur on somewhat consistent timescales, which we can use to test the strength of parametric resonance.
$a(t)$ will typically vary with a frequency comparable to $\Omega_{\textrm{rms}}=\sqrt{\ev{\Omega^2}}$, and as shown in \cite{wzu}, $\ev{\Omega^2}=\frac{n_{\textrm{f}}\Lambda^4}{6\pi} $.
This dependence agrees with \Cref{sect:ccprob}, which gave justification that $T_{00}=\rho_{\textrm{vac}}$ (and thus $\Omega^2$) should scale with $n_{\textrm{f}}\Lambda^4$.
Thus $a$ typically varies with frequency $\sim \sqrt{n_{\textrm{f}}} \Lambda^2$, i.e.\ on a timescale of about $1/(\sqrt{n_{\textrm{f}}}\Lambda^2)$.

\begin{figure} \centering
	\includegraphics[width=0.8\linewidth]{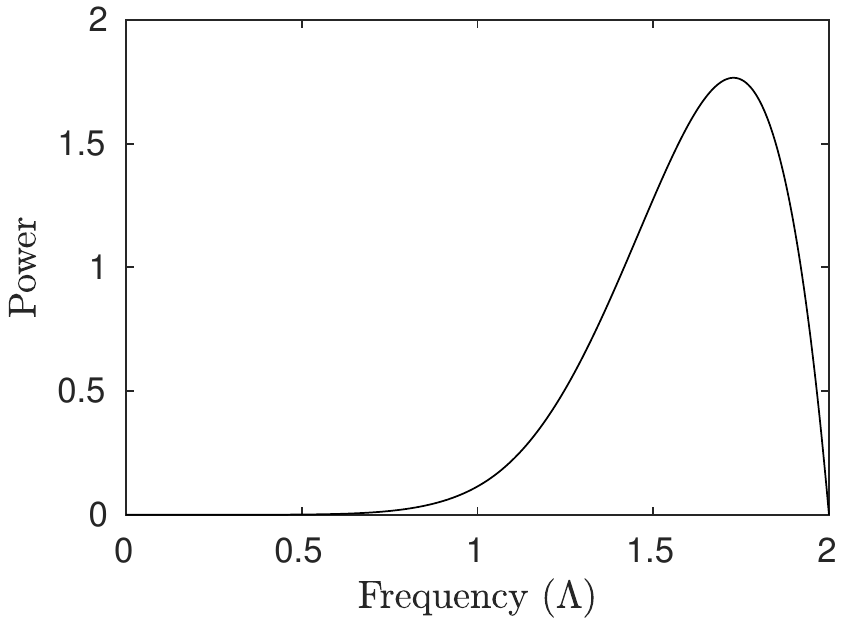}\caption{Normalized power spectrum of $\Omega^2$, showing that small frequencies contribute negligibly, and only those on the order of $\Lambda$ are significant.
	Source: \cite{wzu}}
\label{fig:PowerSpectrum}
\end{figure}

Analysis from \cite{wzu} shows that the power spectrum of $\Omega^2$, on the other hand, is given by \Cref{fig:PowerSpectrum} (independently of the number of fields).
The field amplitude oscillates at all frequencies up to the cutoff, and $\Omega^2$ is proportional to the energy of the vacuum, which scales with the square of the field, so it will oscillate at up to twice the cutoff.
With the average energy of each mode scaling with frequency as $\frac{1}{2} \omega$, we expect that the modes with larger frequencies will dominate as they fluctuate the most violently---with lower frequencies being less significant, as \Cref{fig:PowerSpectrum} confirms.
It follows that the typical timescale for oscillations of $\Omega^2$ (or $\Omega$) will be on the order of $1/\Lambda$.

As mentioned, parametric resonance is strongest when these timescales are similar. 
\change{For the sake of discussing the parametric resonance strength, suppose that $\Omega^2$ only oscillated at a single frequency $\gamma$.
In that case, parametric resonance would be strongest when $r \equiv \frac{2\Omega}{\gamma} \approx 1$, with smaller peaks occuring near higher integers $r\in \N$ \cite{parressummary}.}
Using the results of the previous discussions, $r$ is approximately $2\sqrt{\frac{n_{\textrm{f}}\Lambda^4}{6\pi}}/2\Lambda = \frac{\sqrt{n_{\textrm{f}}}\Lambda}{\sqrt{6\pi}} $, so we expect a peak near $\sqrt{n_{\textrm{f}}}\Lambda \approx \sqrt{6 \pi} \approx 4$ (and weaker peaks at other integer multiples of $\sqrt{6\pi}$).

Away from this ``sweet spot,'' the oscillations in $a$ are typically much faster or slower than the oscillations in $\Omega$ in the limits $\sqrt{n_{\textrm{f}}} \Lambda \to \infty$ and $\sqrt{n_{\textrm{f}}}\Lambda\to 0$, respectively.

For $\sqrt{n_{\textrm{f}}}\Lambda\to 0$ (i.e.\ $\Lambda\to 0$, because $n_{\textrm{f}}$ must be at least one), $\Omega$ oscillates much faster than $a$.
$\Omega$ does not change for long enough to make any one cycle of $a(t)$ significantly different to any other, so $a(t)$ should approach a strictly periodic function, and $H\to 0$ [from \Cref{eq:WzuSolution}].

\begin{figure} \centering
	\includegraphics[width=0.8\linewidth]{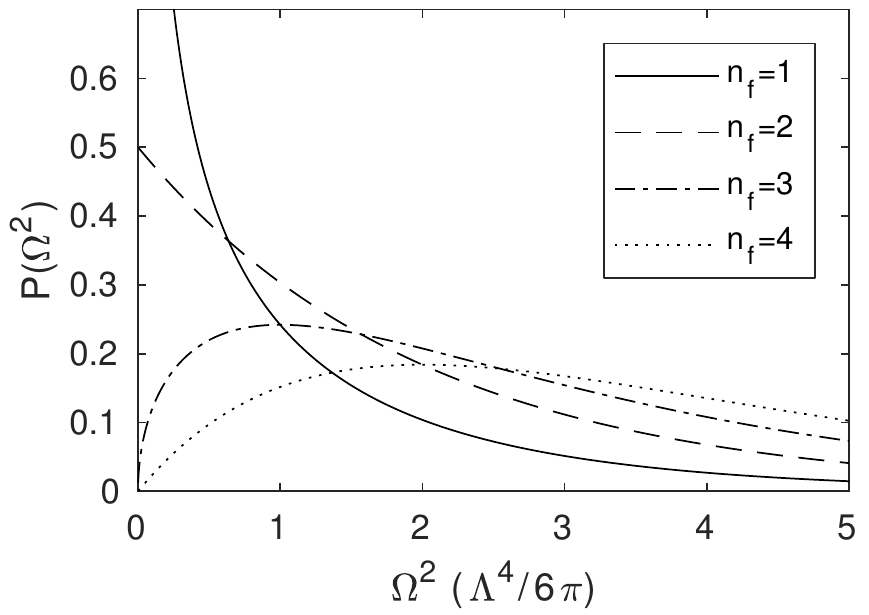}\caption{The probability distribution for $\Omega^2$ at any given time, dependent on the number of particle fields $n_\textrm{f}$.
	With one or two fields, $\Omega^2$ is often arbitrarily small relative to its expectation, but with more fields the fit approaches a Gaussian and the adiabatic limit is more accurate, giving \Cref{eq:WzuH}.}
\label{fig:chi2}
\end{figure}
In the case of $\sqrt{n_{\textrm{f}}}\Lambda\to \infty$, $\Omega$ generally varies much more slowly than $a$.
If the oscillations are consistently slower (known as the adiabatic limit), then a well-known theorem \cite{adiabatic} implies the conservation of the adiabatic invariant (defined as $I(t)=E(t)/\Omega(t)$).
However, although $a(t)$ typically varies on a timescale $1/\Omega_{\textrm{rms}}\sim 1/\sqrt{n_{\textrm{f}}}\Lambda^2$, it can still vary much more slowly if $\Omega$ fluctuates to a very small value.
It becomes important to consider the probability distribution of $\Omega^2(t)$ values, which turns out to follow a $\chi^2_{n_\textrm{f}}$ distribution (a $\chi^2$ distribution with $n_{\textrm{f}}$ degrees of freedom) as shown in \Cref{sect:dist}.
As shown in \Cref{fig:chi2}, $\Omega^2$ will frequently fluctuate to arbitrarily low values with one or two fields.
However, with three or more fields, the chance of $\Omega^2$ fluctuating to a very low value decreases rapidly (exponentially with enough fields), and the timescale of fluctuations in $a(t)$ is more consistently on the order of $1/\sqrt{n_{\textrm{f}}}\Lambda^2$.
%

\begin{figure} \centering
	\includegraphics[width=0.8\linewidth]{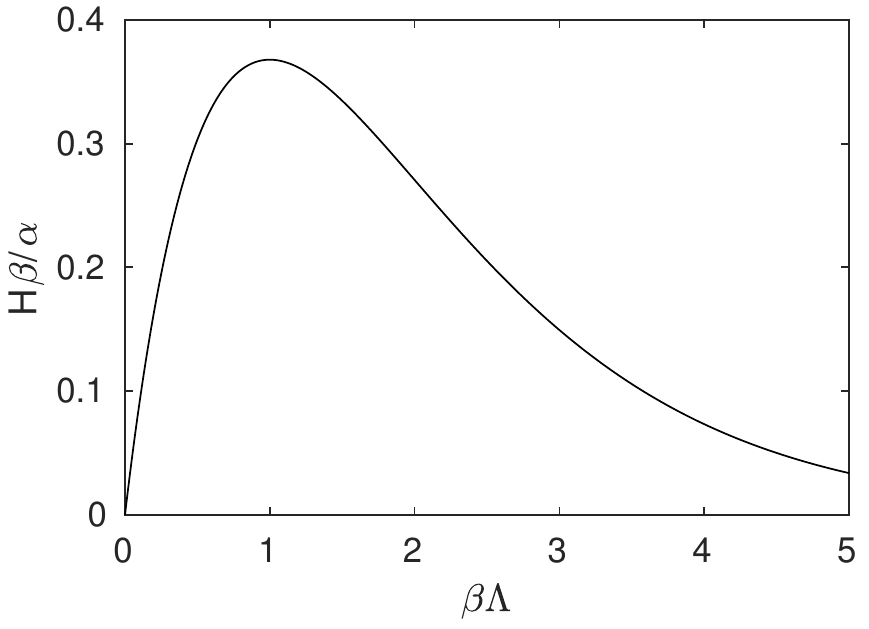}\caption{The form of $H(\Lambda)$ predicted by \cite{wzu}, in the limit of $\Lambda\gg 1$ and shown in normalized units.
	$H$, the expansion rate of the Universe, increases with respect to the high-frequency cutoff $\Lambda$ before reaching a turning point at $\Lambda=\frac{1}{\beta}$, at which it begins to decrease exponentially.}
\label{fig:WzuH}
\end{figure}

It was shown in \cite{wzu} that in the adiabatic limit, the asymptotic dependence of $H$ on $\Lambda$ is:
\begin{align}
	H = \alpha \Lambda e^{-\beta\Lambda},
	\label{eq:WzuH}
\end{align}
where $\alpha$ and $\beta$ are constants.
This relationship is depicted in \Cref{fig:WzuH}.
Although it was stated in \cite{wzu} that this equation is always valid at sufficiently large $\Lambda$, we have seen here that this does not guarantee the adiabatic limit in the cases of $n_{\textrm{f}}=1$ to $3$.
Nonetheless, with more fields, we still expect an exponential decrease of $H$ with respect to $\Lambda$, providing a mechanism for $H^2 \sim 10^{-120}$ as desired.

\begin{figure*}[ht!]
	\centering
	\includegraphics[width=\figwidth]{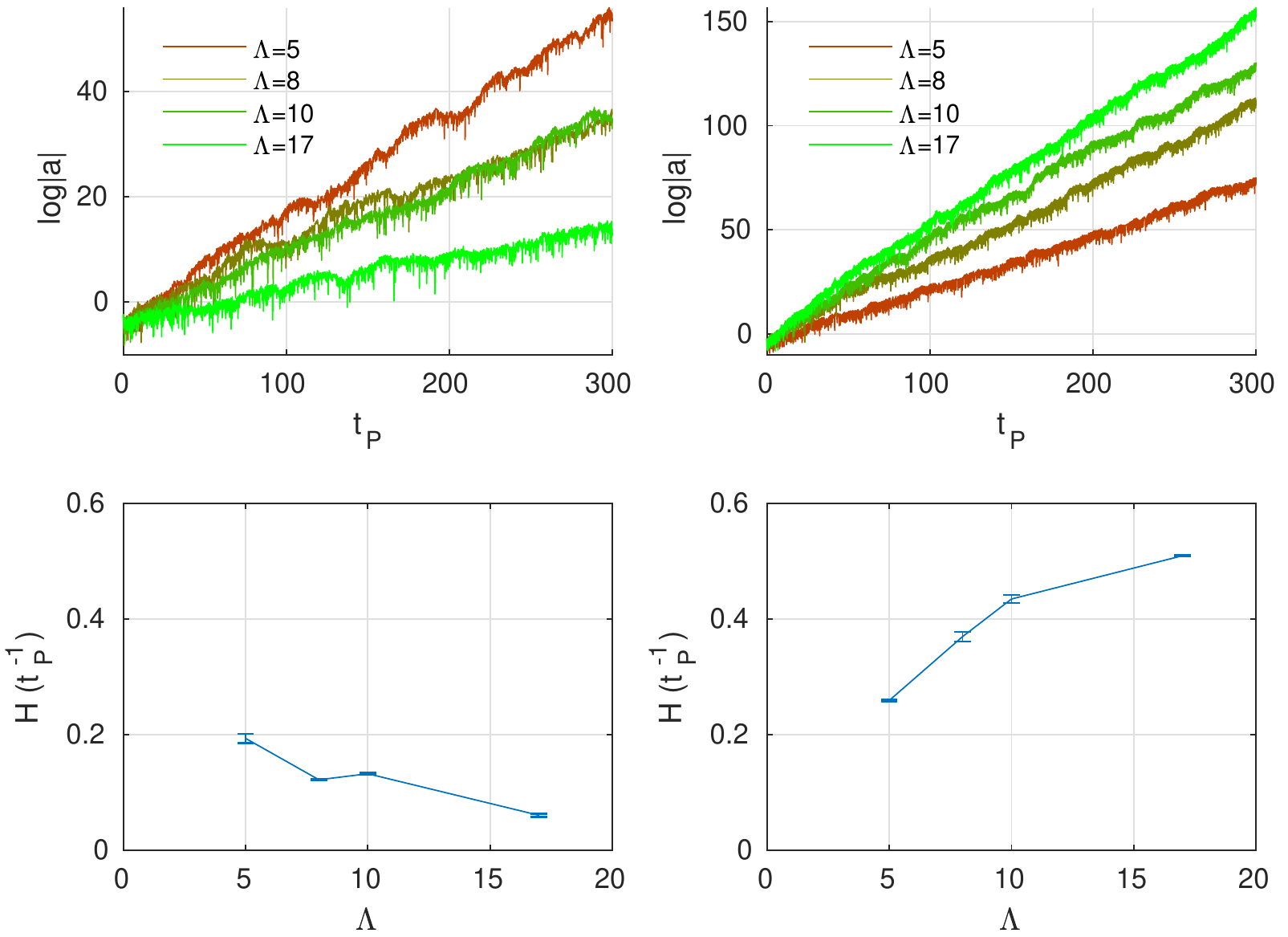}
	\caption{The top two plots display the evolution of the scale factor at the cutoff values $\Lambda$ used in \cite{wzu}.
		The left uses time resolution $t_{\textrm{res}} = 0.15$ (i.e.\ the spacing between evaluations of $\Omega$), which erroneously indicates that $H$ decreases with $\Lambda$ (as shown in the bottom left).
		This is very similar to Fig 5 of \cite{wzu}. 
		On the right, a finer time resolution $t_{\textrm{res}} = 0.01$ is used, showing the correct relationship between $H$ and $\Lambda$ (which persists if time resolution is increased even further).
		This corrected relationship does not exponentially decay to zero as originally claimed, meaning that the cosmological constant problem cannot be resolved by simply taking the cutoff to be $\Lambda\sim 1000 E_P$.
}
	\label{fig:oldtest}
\end{figure*}

\section{Numerical Methods}\label{sect:sims}
In \cite{wzu}, numerical methods were employed to test \Cref{eq:WzuH}, which are also used here.
We will outline the approach used, emphasizing the role of resolution parameters with respect to which our results must converge, before showing detailed convergence tests.

We follow the Wigner-Weyl description of quantum mechanics as used in \cite{wzu} to describe the vacuum energy-momentum tensor, and by extension $\Omega^2$. 
Using this method, we define a pair of coordinates $ x_{\vb k}$ and $p_{\vb k} $ for each mode of the field, indexed by momentum $\vb k$.
These do not represent actual position and momentum coordinates (each mode has well-defined momentum and is completely unlocalized), but instead represent the phase information of the simple harmonic oscillator that describes the mode.
A particular state is represented by a distribution over these variables, $W( \left\{ x_{ \vb k } \right\},\left\{ p_{\vb k} \right\},t )$, where $\left\{ x_{\vb k} \right\}$ denotes the set $\left\{ x_{\vb k_1}, x_{\vb k_2},\ldots \right\}$ with all possible momenta $\vb k$.
Any quantum operator $\hat{A}$ can be represented by a function over these variables, $A\left( \left\{ x_{ \vb k } \right\},\left\{ p_{\vb k} \right\},t \right)$, and its expectation for a state is given by integrating over the state's corresponding distribution:
\begin{align}
	\ev{\hat{A}} = \int_{}^{}\prod_{\vb k}^{}\left( \dd x_{\vb k} \dd p_{\vb k} \right)A\left( \left\{ x_{ \vb k } \right\},\left\{ p_{\vb k} \right\},t \right) W( \left\{ x_{ \vb k } \right\},\left\{ p_{\vb k} \right\},t ).
	\label{eq:WignerExpectation}
\end{align}

In the case of the vacuum state, and using the normalized units from \cite{wzu}, the state distribution is a product of Gaussians:
\begin{align}
	W( \left\{ x_{\vb k} \right\},\left\{ p_{\vb k} \right\},t ) = \frac{1}{\pi} \prod_{\vb k}^{} e^{-p_{\vb k}^2 - x_{\vb k}^2 } .
	\label{eq:WignerVacuum}
\end{align}
It is quite difficult to numerically perform the integral in \Cref{eq:WignerExpectation}, because there can be a very large number of modes (i.e.\ many values of $\vb k$), meaning that this is an integral over many dimensions.
Fortunately, there is an alternative method.
If we randomly sample $ \left\{ x_{\vb k} \right\}$ and $\left\{ p_{\vb k} \right\} $ from the distribution given by \Cref{eq:WignerVacuum}, and then perform an average over the resultant solutions of $A\left( \left\{ x_{\vb k} \right\},\left\{ p_{\vb k} \right\},t \right)$, the different regions of phase space will be appropriately weighted by their likelihood of being chosen.
As the number of randomly sampled points $N$ increases, the resultant value will converge to the true result from \Cref{eq:WignerExpectation}.

Now, we can choose an operator $A$ to evaluate.
We wish to examine what happens on average to \change{$a(t,\vb x)$} in \Cref{eq:WzuDynamics} at a single point in space over time. 
This means we must describe $\Omega^2$ as a function $\Omega^2\left( \left\{ x_{\vb k} \right\},\left\{ p_{\vb k} \right\},t \right)$ using the above formulation, evaluate it for $N$ different choices for the sets of random numbers $ \left\{ x_{\vb k} \right\}$ and $\left\{ p_{\vb k} \right\} $, solve for $a\left( \left\{ x_{\vb k} \right\},\left\{ p_{\vb k} \right\},t \right)$, and then average the results to determine $\ev{a(t)}$.
Alternatively, one could apply \Cref{eq:WignerExpectation} to $H$ instead of $a(t)$, to compute the expectation value $\ev{H} = \langle \frac{\dot{a}}{a} \rangle$.
We will discuss this further shortly.

The expression for $\Omega^2$ in terms of quantum operators contains contributions from the infinite continuum of allowed momenta values $\vb k$.
Even if a cutoff energy (or equivalently, cutoff frequency) $\Lambda$ is applied, there will still be continuously infinitely many modes to consider.
To make it suitable for numerical calculation then, we need to discretize it, which can be done by considering a cube of width $L$ in physical space, and restricting the allowed modes of our field to be only harmonic modes of the box.
$L$ is another parameter with respect to which our results should converge to a consistent, physical solution, specifically in the limit $L \to \infty$.
Harmonic modes in this box are proportional to $\sin(\frac{n_x 2 \pi x}{L})\sin(\frac{n_y 2 \pi y}{L})\sin(\frac{n_z 2 \pi z}{L})$, for some set of integers $(n_x, n_y,n_z)$ (each of which can be positive or negative) that we call $\vb  n$.
The corresponding frequency is $\omega=\frac{ 2 \pi|\vb n|}{L}$, so we can translate the cutoff $\omega\le \Lambda$ to a cutoff on $\vb  n$ by $n_{\textrm{max}} = \frac{L\Lambda}{2 \pi} $.
In \cite{wzu}, this cutoff was applied to each component, i.e.\ $n_{x,y,z}\le n_{\textrm{max}}$.
Whereas this would signify a cube of allowed modes in momentum space, with side length $2\Lambda$ and maximum frequency $\sqrt{3}\Lambda$, we instead apply the cutoff as a sphere in momentum space of radius $\Lambda$ by choosing modes with $|\vb n| < n_{\textrm{max}}$.
Now, our sets $\left\{ x_{\vb k} \right\}$ and $\left\{ p_{\vb k} \right\}$ are labeled as $\left\{ x_{\vb n} \right\}$ and $\left\{ p_{\vb n} \right\}$, and they each contain one random number for every value of $\vb n=(n_x,n_y,n_z)$ such that $|\vb n|<n_{\textrm{max}}$.

For a particular cutoff method, \cite{wzu} shows that we can write $\Omega^2$ for a single massless scalar field as 
\begin{align}
	\Omega^2\left( \left\{ x_{\vb n} \right\},\left\{ p_{\vb n} \right\},t \right) &= \left[ \sum_{\vb n} \sqrt{n}\left( x_{\vb n} \sin\left( n t \right) - p_{\vb n} \cos \left( nt \right) \right) \right]^2.
	\label{eq:WignerEnergy}
\end{align}
The above just describes the process for a single massless scalar field.
To incorporate more, it is repeated for each individual $\Omega^2_j$ and then the total is computed as $\Omega^2 = \sum_{j=1}^{n_\textrm{f}}  \Omega^2_j$.

After randomly sampling $ \left\{ x_{\vb n} \right\}$ and $\left\{ p_{\vb n} \right\}$ values for each field and computing $\Omega^2$ at a number of points in time (with spacing $t_{\textrm{res}}$ up to a maximum $t_f$, two more parameters to test for convergence), the differential equation in \Cref{eq:WzuDynamics} is solved for $a (t)$ by interpolating $\Omega^2$.
Doing this $N$ times, either $\frac{\langle \dot{a}(t) \rangle}{\langle a(t) \rangle}$ or $\ev{ \frac{ \dot{a}(t) }{ a(t) }}$ can be determined from \Cref{eq:WignerExpectation}, and then a time average taken to find $H$.

\subsection{Correction to previous results}
When implementing these methods, we found substantially different results to those of \cite{wzu}, as shown in Figure~\ref{fig:oldtest}.
Investigation showed that due to a combination of factors, the original calculations did not properly capture the dynamics of $\Omega$ at fine enough timescales. 
When computing $\Omega^2$ from \Cref{eq:WignerEnergy}, the spacing $t_{\textrm{res}}$ must be at least as small as the timescale on which we expect oscillations in $\Omega$ to occur; otherwise, the numeric description of $\Omega$ will not display the high-frequency behavior of the actual function (which is particularly significant for parametric resonance, as discussed in the previous section).
This was the key problem with the original calculations: $t_{\textrm{res}}$ was too large to have sufficiently converged.
Furthermore, it was not made finer for higher $\Lambda$, so more and more of the significant short-timescale behavior was lost for higher $\Lambda$.
Finally, a simple linear interpolation method was used rather than a smooth method when determining $\Omega$, which exacerbated the resolution problem (see Fig.~\ref{fig:trestest}).
The impact of these differences on the resultant scale factor is shown in Figure~\ref{fig:oldtest}, which shows that the relationship between $H$ and $\Lambda$ is drastically affected.
We discuss the implications of these changed results in \Cref{sect:results}.

\subsection{Convergence tests}
Having found that the discrepancy between our results and those of \cite{wzu} was due to different time resolution parameter values, we sought to validate that no other resolution parameters were being overlooked.
Let us recap the roles of the relevant parameters: we generate $N$ instances of the random sets $ \left\{ x_{\vb n} \right\}$ and $\left\{ p_{\vb n} \right\} $, which each contain a random number for every integer vector $\vb n$ with magnitude $|\vb n|<\frac{L\Lambda}{2\pi}$, where $L$ is the size of the box and $\Lambda$ is the maximum frequency permitted.
These produce a frequency function $\Omega(t)$ using \Cref{eq:WignerEnergy}, which we evaluate at evenly spaced points between $t=0$ and $t=t_f$, with spacing $t_{\textrm{res}}$. 
Then, we interpolate between those points to solve the differential equation $\ddot a(t) = - \Omega^2(t) a(t)$ with initial conditions $(a(0),\dot{a}(0)) = (1,0)$.
Averaging over the $N$ different samples, we then determine our average expansion rate $H$.

There are five variables here with respect to which our results should converge: the box width $L$, the final time $t_f$, the time resolution $t_{\textrm{res}}$, the number of samples being averaged $N$, and the relative tolerance of the ODE solver, which we will denote $\varepsilon$.
There are also several qualitative choices which may affect the results: whether the cutoff should be implemented as a cube or sphere in momentum space, how to interpolate $\Omega$ when solving the differential equation \Cref{eq:WzuDynamics}, and how to determine $H$ given the solution $a(t)$.
We present discussions for each of these in \Cref{sect:convergence}, except for that of $H$ which we present now.

\subsection{New method of determining $H$}
In our new tests, we made a number of changes to the implementation to improve the efficiency.
Most of these did not represent physical differences in what was being computed, but one exception is the method of determining $H$.
Physically, the Hubble Constant  $H$ is defined as $H=\frac{v}{d}$, in which $v$ is the radial outwards velocity of a remote astronomical object and $d$ is its distance to the earth.
To determine $v$ and $d$ one needs to measure at least two properties, the redshift of a galaxy as well as an independent measure of its distance, such as the luminosity of a type Ia supernova, or the length of a standard ruler.
So in principle, we need to study the behavior of a long wave photon field propagating on our wildly fluctuating metric to determine $H$.
Technically, we need to solve the wave equation in our inhomogeneous ``FLRW'' metric (23) in \cite{wzu}:
\begin{equation}\label{wave equation}
\nabla^{\mu}\nabla_{\mu}\phi=\frac{1}{\sqrt{-g}}\partial_{\mu}\left(\sqrt{-g}g^{\mu\nu}\partial_{\nu}\phi\right)=0.
\end{equation}
Unfortunately, this is a nontrivial calculation which is beyond the scope of this article.  The usual definition of $H$ in cosmology, i.e. $H=\dot{a}/a$ depends on the validity of the homogenous FLRW metric. For the generalized inhomoegenous FLRW metric (23) in \cite{wzu}, we can have a similar definition as
\begin{equation}\label{hubble expansion definition}
H(t)=\frac{\dot{L}}{L}=\frac{\int_{\mathbf{x}_1}^{\mathbf{x}_2}\frac{\dot{a}}{a}(t, \mathbf{x})\sqrt{a^2(t, \mathbf{x})}dl}{\int_{\mathbf{x}_1}^{\mathbf{x}_2}\sqrt{a^2(t, \mathbf{x})}dl}.
\end{equation}
The macroscopic Hubble constant is acquired by taking both the spatial and temporal average on both sides of  \eqref{hubble expansion definition}, as well as the average in the phase space using the Wigner-Weyl representation to get its quantum expectation value. 
\begin{figure}[ht!]
	\centering
	\includegraphics[width=\linewidth]{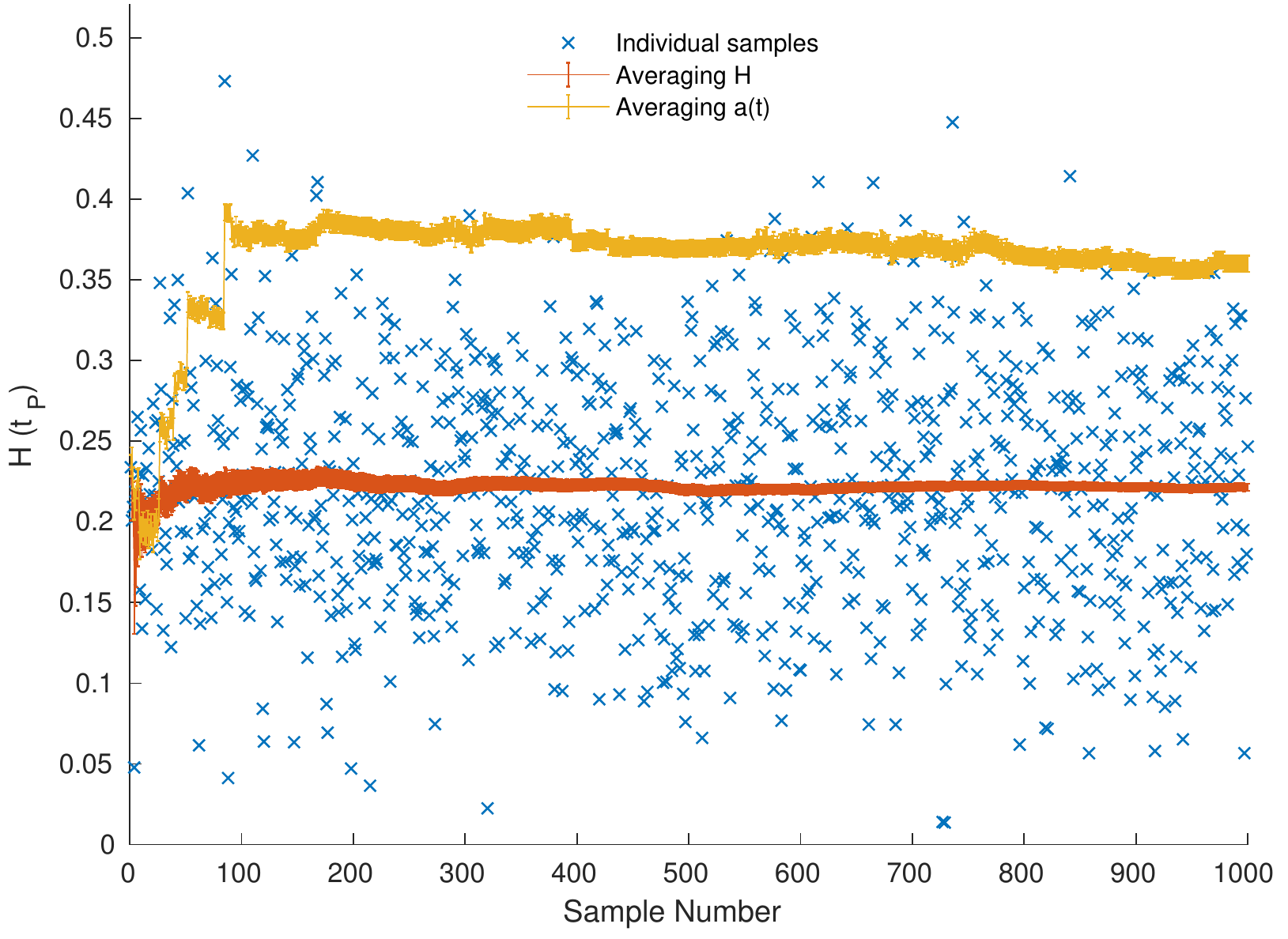}
	\caption{Here (with $\Lambda=5$ and $L=10$), we compare methods of determining $H$: that of averaging $\ev{a(t)}$ first, as done in \cite{wzu}, and our approach which was to average $H$ directly as $H=\ev{\frac{\dot{a}(t)}{a(t)}}$.
}
	\label{fig:htest}
\end{figure}

In \cite{wzu}, the expectation value of the scale factor $\langle a(t) \rangle$ is determined first by Wigner-Weyl formulation, and then $H$ is calculated as the time average of $H (t ) =\frac{\dot{\langle a(t) \rangle}}{\langle a(t) \rangle}$.
However, we can also change the sequence of averaging and directly compute the expectation value of H, by using \Cref{eq:WignerExpectation} to calculate $H (t) =\langle\frac{\dot{a(t)}}{a(t)}\rangle$.
In this way, we actually define $H$ as the time average
\begin{equation}
	\label{def H} H=\overline{\left(\frac{\dot{a}(t)}{a(t)}\right)}.
\end{equation} 
It is more physical compared to the original case in \cite{wzu} since the scale factor $a(t)$ (being an arbitrary distance scale) is less fundamental than the actual distance between objects.
Given that $\dot{a}/a$ is also equivalent to $\frac{d\log|a|}{dt}$, this choice means that an average is computed in logarithmic space with respect to a, rather than linear space.
Not only does this method lead to a different value for $H$ which is physically better justified, but computation of this value is also much easier and more stable, as shown in \Cref{fig:htest}.
This is because the linear method is heavily biased towards the samples with the largest $H$, resulting in high sensitivity to the occasional outlier, so it has much slower convergence.
Instead, the logarithmic method (averaging $H$) quickly converges to a consistent result about which the distribution of individual samples appears to be roughly symmetric.
 
As we mentioned before,  our new definition of $H$, which is based on the distance definition \eqref{hubble expansion definition}, is not necessarily equivalent to the observed Hubble constant in astronomy.
The observed Hubble constant should be acquired by solving \Cref{wave equation} for the actual redshift and intensity damping of a macroscopic light signal.
However, we believe that the calculation of H based on \Cref{def H} can still provide useful insight about how the actual Hubble Constant behaves in this metric.

\section{New Results}\label{sect:results}
\begin{figure*}[ht!]
	\centering
	\includegraphics[width=1.25\figwidth]{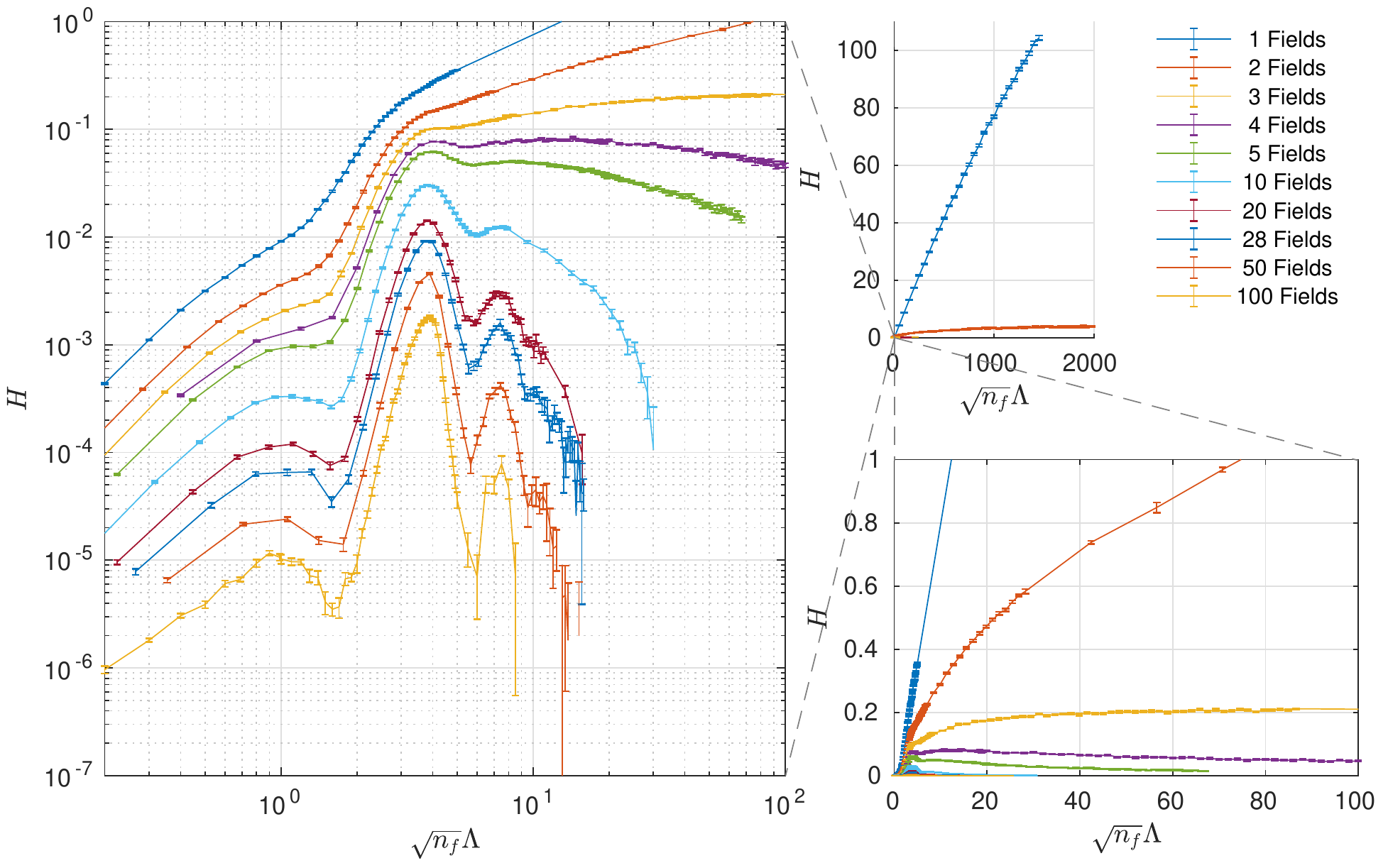}
	\caption{Here we see the relationship between $H$ and $\Lambda$ with $n_{\textrm{f}}$ fields. 
		On the left it is shown as a log-log plot, so that all the regimes can be seen at once.
		On the right, the top shows more clearly the linear increase for one field and the logarithmic increase for two (in linear space).
		The bottom right shows the beginning of the turning point for three fields, and exponential decay for several higher $n_{\textrm{f}}$ cases (which \Cref{eq:WzuH} predicted for all numbers of fields).
		Note that we have used $\sqrt{n_{\textrm{f}}} \Lambda$ on the $x$-axis for each of these rather than just $\Lambda$, because this is the term on which the adiabatic limit depends, as well as the resonances described in \Cref{sect:stochastic}.
}
	\label{fig:morefieldsnotheory}
\end{figure*}

\begin{figure*}[ht!]
	\centering
	\includegraphics[width=\figwidth]{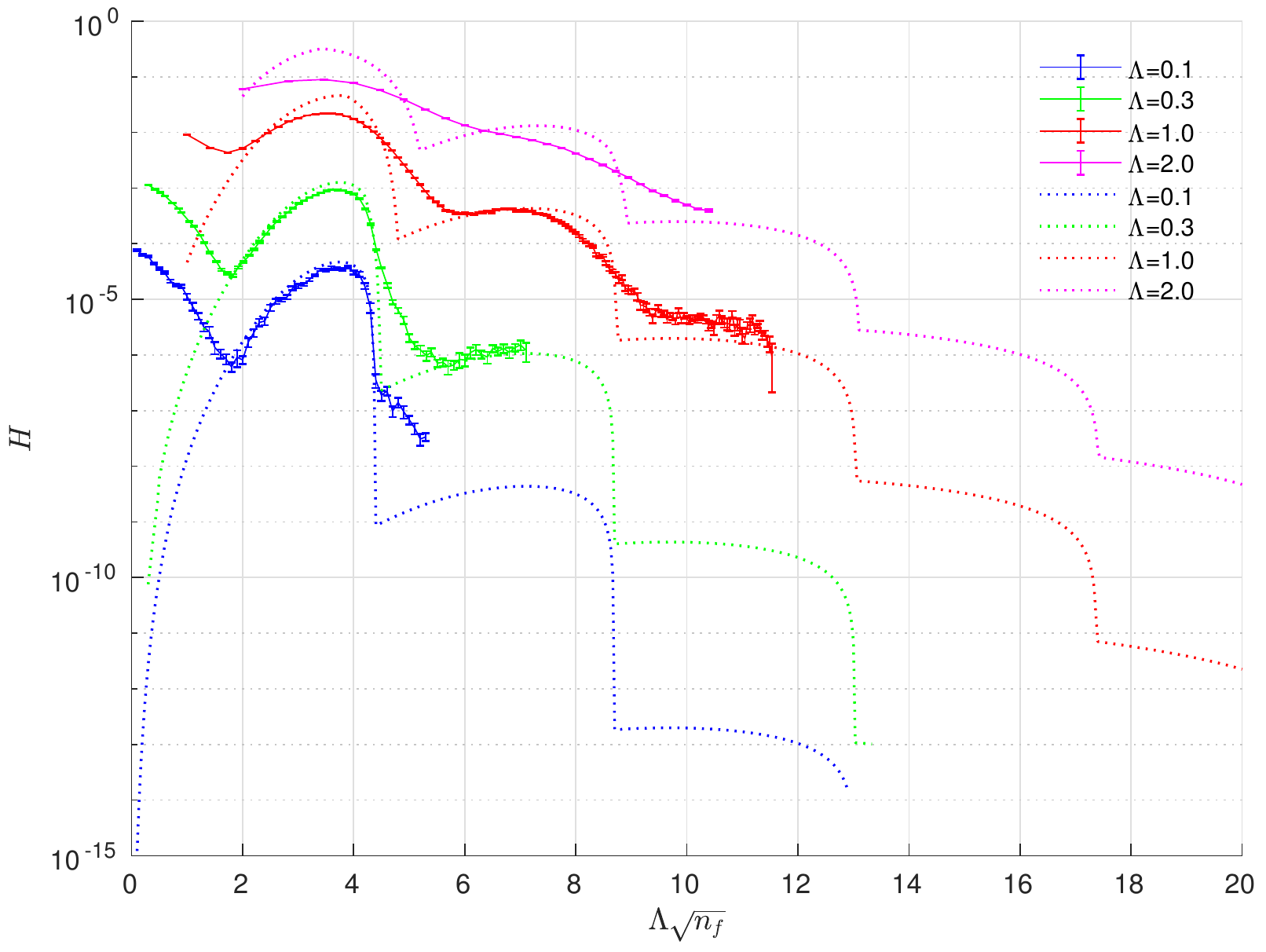}
	\caption{Here we see $H$ against $n_{\textrm{f}}$ for fixed cutoff, which exhibits a decreasing step-like relationship.
		The dotted lines represent the results of an approximation described in the text which allows an analytical prediction of the behavior of $H$.
		We see that although they do not precisely match the results, these approximations do predict the existence and approximate size of the periodic steps downward, and the relationship between the two curves.
		For low numbers of fields, the error is dramatic, but the fit improves as $n_{\textrm{f}}$ increases, because the approximation that $\Omega^2$ is roughly constant vastly improves.
}
	\label{fig:fixedcutoff}
\end{figure*}

\begin{figure*}[ht!]
	\centering
	\includegraphics[width=\figwidth]{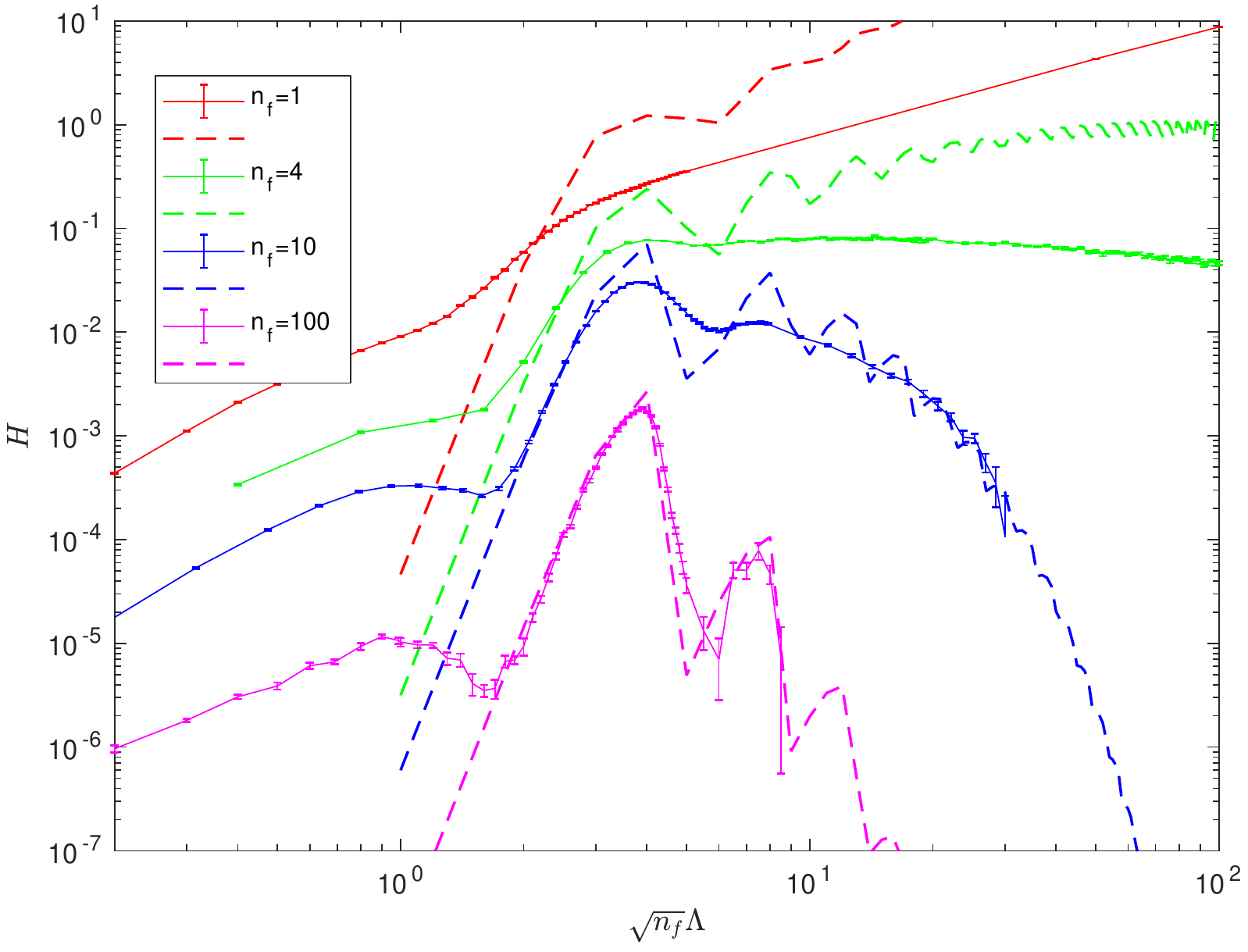}
	\caption{A similar model to that of \Cref{fig:fixedcutoff} is shown in dashed lines, and some of the simulation results from \Cref{fig:morefieldsnotheory} are shown in solid lines.
		Again, the results are not matched precisely, but the fit is quite good as $n_{\textrm{f}}$ increases, because the approximation that only one resonance contributes significantly becomes vastly more accurate.
		At low $\Lambda$, resonances no longer occur near the mean of $\Omega^2$, and the approximation also worsens.
}
	\label{fig:morefields}
\end{figure*}

In Figure~\ref{fig:morefieldsnotheory}, we see that the relationship between $\Lambda$ and $H$ is quite complex, with the behavior of the curve depending significantly on the number of fields.
\Cref{fig:fixedcutoff} then shows the relationship between $H$ and $n_{\textrm{f}}$ for several choices of fixed $\Lambda$, to examine what happens to $H$ if we enforce an approximate Planck cutoff ($\Lambda\sim 1$) and then vary the number of fields.
We will first compare the findings of \Cref{fig:morefieldsnotheory} to the proposed relationship \Cref{eq:WzuH}, and check that the limiting behaviors predicted in \Cref{sect:stochastic} are satisfied.
After this, we will introduce a model which captures important features of the behavior of $H$ vs $n_{\textrm{f}}$, shown in \Cref{fig:morefields,fig:fixedcutoff}, and use it to estimate the number of scalar fields required to achieve $H\sim 10^{-60}$, such that $\lambda_{\textrm{eff}}\sim 10^{-120}$.

\begin{figure}[ht!] \centering
	\includegraphics[width=\linewidth]{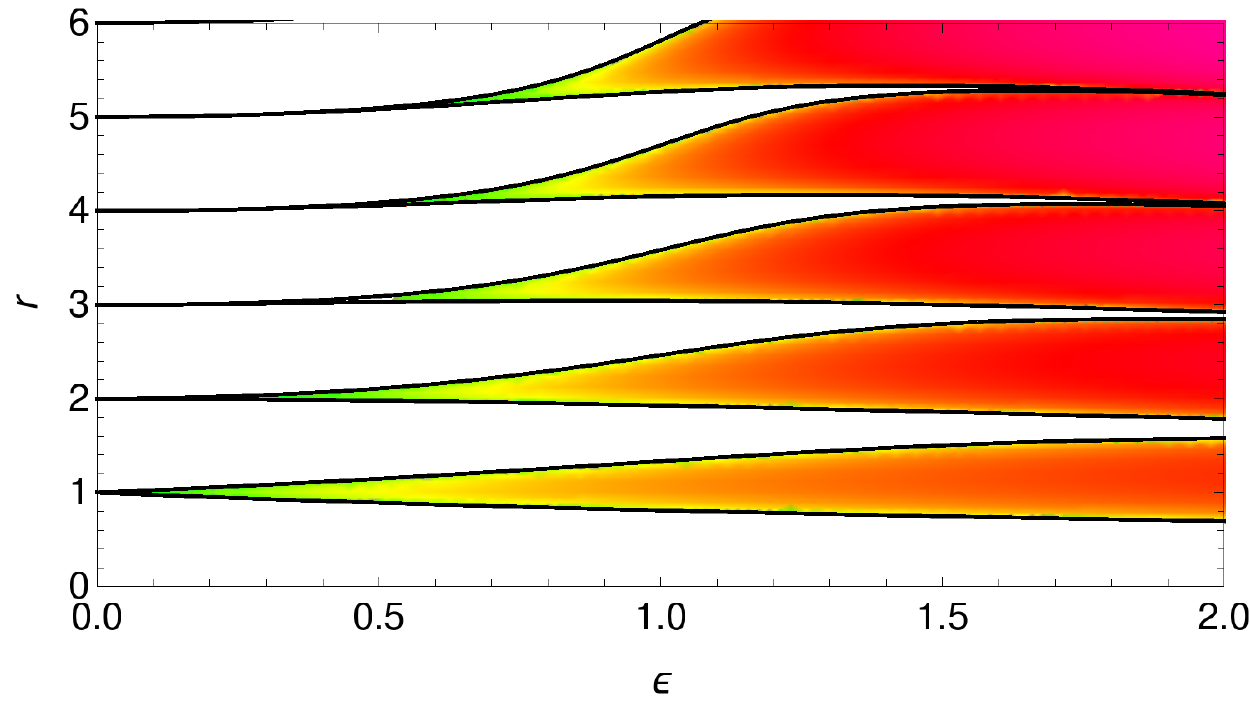}
	\includegraphics[width=\linewidth]{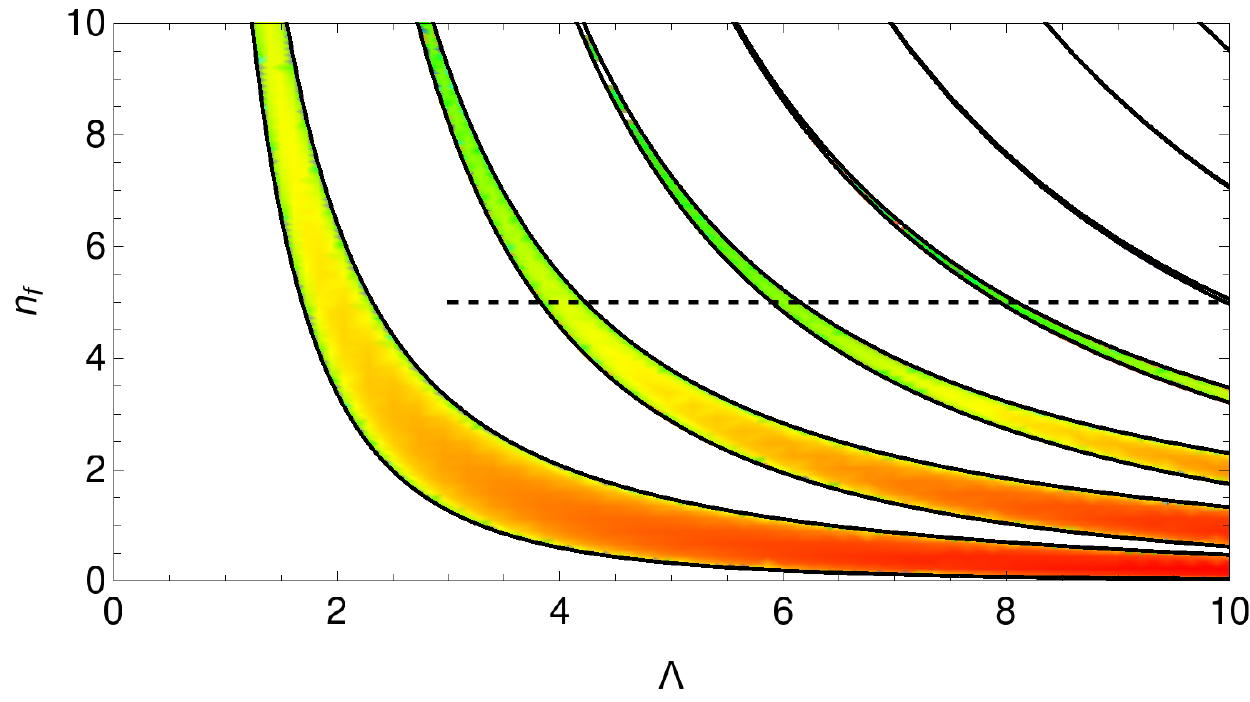}
	\caption{
		The stability regimes of the Mathieu equation, a prototypical example of parametric resonance.
		Colored regions indicate instability, in which $H>0$, with larger $H$ represented by redder colors.
		For a given $n_{\textrm{f}}$ and $\Lambda$, the highest frequencies at which the vacuum oscillates correspond to a resonance point given by the lower graph.
		It also oscillates at all frequencies below that value, which here means all points directly to the right of that point.
		Even if the highest frequency does not lie in any of the resonance bands and excite resonance directly, as is the case for $(\Lambda,n_{\textrm{f}})=(5,3)$, the fluctuations at lower frequencies can excite resonances directly to the right of the point, as indicated by the dashed black line.
		In this example, the second band dominates the growth, because no vacuum oscillations occur in the stronger first band.
}
\label{fig:stability}
\end{figure}

As explained in \Cref{sect:stochastic}, we expect that $\lim_{ \Lambda\to 0 } H=0$, which seems to hold in all cases.
\Cref{eq:WzuH} also predicts that, for a given number of fields and at large-enough $\Lambda$, there will be an exponentially decreasing relationship $H\sim e^{-\beta\Lambda}$.
On the left, in log-log space, such a relationship appears as $\log(H) \sim - \beta e^{\log(\Lambda)}$, which upon inspection, seems to match the large-$\Lambda$ behavior for $n_{\textrm{f}}\ge 4$.
As discussed in \Cref{sect:stochastic}, this relied upon the adiabatic theorem which is only valid when $n_{\textrm{f}} \gtrsim 3$.
Indeed, this trend does not seem to hold for $n_{\textrm{f}}<4$, (the behavior for $n_{\textrm{f}}=1$ at large $\Lambda$ appears to be linear, and for $n_{\textrm{f}}=2$ and $3$ it appears to be logarithmic).
While there may be some turnaround at higher $\Lambda$ (and the linear behavior of $n_{\textrm{f}}=1$ may become logarithmic at some higher $\Lambda$), this does not occur in the regime checked, which is up to $\Lambda\approx 1500$.


Note that there are peaks corresponding to those predicted in \Cref{sect:stochastic}, near $\sqrt{n_{\textrm{f}}}\Lambda \approx 4$, and a weaker one near $8$.
These resonances draw a direct parallel with the behavior of the Mathieu equation, a simple prototypical example of parametric resonance \cite{mathieucomputation,landau}.
the Mathieu equation takes the following simplified form, in which the fluctuations to $\Omega^2$ are strictly periodic with constant amplitude:
\begin{align}
	\ddot{a}(t) = -\Omega^2_0 \left( 1+ \epsilon \cos(\gamma t) \right) a(t).
	\label{eq:Mathieu}
\end{align}

Solutions of the Mathieu equation take the following form, similar to the right-hand side of \Cref{eq:WzuSolution}:
\begin{align}
	a(t) = e^{Ht}P(t)\text{, where $P(t)=P\left(t+\frac{2\pi}{\gamma}\right)$}.
	\label{eq:MathieuSolution}
\end{align}
$H$ can either be real (an unstable solution with exponentially growing solutions) or imaginary (representing stable quasiperiodic solutions with no long-term growth or decay).
The stable and unstable regions depend on $\epsilon$ and $r = \frac{2\Omega_0}{\gamma}$, as shown in \Cref{fig:stability}.
Although no closed-form expressions exist, there are efficient methods of computing both the region boundaries and the magnitude of the exponents \cite{mathieucomputation}.

A novel idea of the current work is to approximate \Cref{eq:WzuDynamics} using the above form, to obtain an approximate model in terms of the simpler, better-understood Mathieu equation.
In \Cref{sect:mathieu}, we find that a sensible set of choices for this approximation is to use $\Omega_0^2 = \ev{\Omega^2} = \frac{n_{\textrm{f}}\Lambda^4}{6\pi}$, $\epsilon=\sqrt{\Var(\Omega^2)}/\ev{\Omega^2} = \sqrt{2/n_{\textrm{f}}}$, and $\gamma$ taking on a range from $0$ to $2\Lambda$ according to \Cref{fig:PowerSpectrum}.
With fixed $\varepsilon$ and $r$ varying as $r=\frac{2\Omega_0}{\gamma}= \frac{2\Lambda^2}{\gamma} \sqrt{\frac{n_{\textrm{f}}}{6\pi}}$, each choice of $n_{\textrm{f}}$ and $\Lambda$ excites a range of resonances from $\gamma=0$ to $\gamma=2\Lambda$ as indicated by the dashed line in \Cref{fig:stability}.
Our approximation is to select out the $\gamma$ with the most significant parametric resonance effect, weighted by the strength it oscillates at according to \Cref{fig:PowerSpectrum}.

Using these methods, we obtain the dotted lines shown in \Cref{fig:fixedcutoff}, which capture many of the key properties (e.g.\ existence and size of the ``steps'' that arise as a result of resonance).
This method explains the steplike behavior of \Cref{fig:fixedcutoff}, because these ``steps'' occur when a resonance band leaves the region of allowed $\gamma$ (e.g.\ when the dashed black line in \Cref{fig:stability} moves high up enough that it does not cover the second band).
These methods can also explain why we see divergence as $\Lambda$ increases for $n_{\textrm{f}}=1$ through $3$.
Looking at the top of \Cref{fig:stability}, one and two fields correspond to $\varepsilon>1$, and in this region the higher-order bands (further from the origin) have a larger amplitude. 
In these cases, as $\Lambda$ increases and the ratio between frequencies of oscillation for $a$ and $\Omega$ increases, the parametric resonance effect gets stronger and $H$ diverges.
At lower numbers of fields, $\varepsilon$ decreases and the trend reverses: on the left of \Cref{fig:stability}, increasing $r$ leads to exponentially weaker resonance, and $H\to 0$.
At high $\Lambda$, the model predicts logarithmic divergence for $1\leq n_{\textrm{f}} \leq 3$, and an asymptotically uniform $H$ for $n_{\textrm{f}}=4$.

In some regions this model clearly does not fit as well as others. 
It approximates $\Omega^2$ as only oscillating at one frequency, and simplifies $a$ by ignoring any squared frequencies outside of the range $\Omega^2 \pm \varepsilon$.
The latter approximation explains why the model fails at low $\Lambda$ and $n_{\textrm{f}}$ in \Cref{fig:morefields}, and for low $n_{\textrm{f}}$ when $\Lambda=0.1$ in \Cref{fig:fixedcutoff}: in these regimes, none of the frequencies that $a$ oscillates at excite resonances, and it is in fact oscillations outside this range which drive the dominant resonances.
On the other hand, when the number of fields increases, the power spectrum for $a$ (see \Cref{fig:chi2}) becomes much narrower, and the Mathieu model is a better description.
This explains the very tight fit for $n_{\textrm{f}}=100$ in \Cref{fig:morefields}.
The approximation that $\Omega^2$ oscillates only at one frequency fails at low fields for the same reason that the adiabatic limit does: in these cases, the distribution of $\Omega^2$ values is too broad (see \Cref{fig:chi2}).
This is why the model does not fit as well for $1$ and $4$ fields in \Cref{fig:morefields} at high $\Lambda$.

The advantage of this model is that compared to the full simulations, it is much easier to calculate for small $H$.
Even though these methods are still restricted by machine precision to $H\sim 10^{-16}$, the trends are consistent and can be extended all the way down to $H=10^{-60}$ so that we can test what cutoff and number of fields would be required to match observation.
If the trend for $\Lambda=1$ continues as shown, then $H=10^{-60}$ will be achieved with $n_{\textrm{f}} \approx 6000$.
Similarly, extending the $n_{\textrm{f}}=28$ line (because $28$ is the number of bosonic field components in the standard model), we get $H=10^{-60}$ (i.e.\ we match observation) when $\Lambda\approx 40$ (i.e.\ cutoff at $40$ times the Planck energy).

\section{Discussion and Conclusion}\label{sect:conclusion}
We will now review the typical assumptions that are made in the usual formulation of the cosmological constant problem (which we refer to as the ``traditional approach''), in order to provide a framework with which we can discuss the significance of our new findings.
In \Cref{sect:ccprob}, we described the problem using a simple case with a single scalar field, but its conclusions hold in a much broader range of contexts.
We summarize the key assumptions (as relating to calculations of vacuum energy) before discussing them in further detail:
\renewcommand{\labelenumii}{\theenumi$^\prime$}
\renewcommand{\theenumii}{$^\prime$}
\renewcommand{\labelenumi}{\theenumi}
\setlength{\leftmarginii}{0ex}
\begin{description}
	\item [Traditional assumptions]
\end{description}
\begin{enumerate}
	\item The total effective cosmological constant $\lambda_{\textrm{eff}}$ is on at least the order of magnitude of the vacuum energy density generated by zero-point fluctuations of particle fields. \label{item:lambda}
	\item QFT is an effective field theory description of a more fundamental, discrete theory, which becomes significant at some high-energy scale $\Lambda$.\label{item:eft}
	\item The vacuum energy-momentum tensor is Lorentz invariant.\label{item:lv}
	\item The Moller-Rosenfeld approach to semiclassical gravity (using an expectation value for the energy-momentum tensor) is sound.\label{item:ev}
		\change{\item The Einstein equations for the homogeneous Friedmann-Robertson-Walker metric accurately describes the large-scale evolution of the Universe.\label{item:hom}}
\end{enumerate}
With these assumptions, one arrives at the usual value of $\lambda_{\textrm{eff}}\sim 1 \sim 10^{120}\lambda_{\textrm{obs}}$. 
However, it has been noted \cite{martin,covariant1,covariant2,covariant3} that there is an inconsistency between \Cref{item:lv,item:eft}: the vacuum state cannot be Lorentz invariant if modes are ignored above some high-energy cutoff $\Lambda$, because a mode that is high energy in one reference frame will be low energy in another appropriately boosted frame.

In the new approach proposed by \cite{wzu}, \Cref{item:lv} is not used and this contradiction is avoided. 
Also, \Cref{item:ev,item:hom} is modified, which we denote as \Cref{item:ev2,item:inhom} respectively, and the simple toy model also introduces \Cref{item:field}:
\subsection*{Modified assumptions}
\begin{enumerate}
				\setcounter{enumi}{3}
					\item[] \stepcounter{enumi}
\begin{enumerate}
	\item The \textbf{semiclassical stochastic approach} to gravity (using a stochastic field for the energy-momentum tensor) is sound.\label{item:ev2}
\end{enumerate}
					\item[] \stepcounter{enumi}
\begin{enumerate}
\change{\item The temporal Einstein equation for the simple inhomogeneous metric \Cref{eq:InhomMetric} is a reasonable approximation to the dynamics of the Universe.\label{item:inhom}}
\end{enumerate}
	\item The Universe can be effectively modeled by a single massless scalar field.\label{item:field}
\end{enumerate}
\subsection{Different contributions to $\lambda_{\textrm{eff}}$}
\Cref{item:lambda} is well justified in the case of the traditional problem, because the contribution from zero-point fluctuations is on the order of $1$ in Planck units and no other known contributions are as large \cite{martin}---thus, assuming no significant cancellation of terms (e.g.\ fine tuning of the bare cosmological constant $\lambda$), the total $\lambda_{\textrm{eff}}$ should be at least on the order of the largest contribution.
In the case of the new approach introduced in \cite{wzu} and used here, this assumption is also reasonable: any other contributions would also presumably fluctuate and result in similar effects to what we have found here.

\subsection{Effective field theory and Lorentz invariance}
To prevent the vacuum energy density from diverging, the traditional approach also assumes that performing a high-energy cutoff is acceptable.
This type of regularization is a common step in renormalization procedures, which aim to eventually arrive at a physical, cutoff-independent result.
However, in the case of the vacuum energy density, the result is inherently cutoff dependent, scaling quartically with the cutoff.

This is acceptable under the philosophy of \Cref{item:eft}, which treats QFT as a low-energy effective field theory and not a fundamental theory.
This approach draws parallels with the case of the ultraviolet catastrophe: the equipartition theorem (a key feature of classical physics) made a rapidly divergent prediction when high-energy modes were considered, but a new high-energy theory (quantum mechanics) resolved this problem, and showed classical mechanics to be only an effective low-energy theory.
Similarly, it is presumed here that a high-energy discrete theory would not display the zero-point fluctuations that are characteristic of QFT, and hence that the divergence caused by oscillations above the corresponding cutoff frequency is unphysical.
In this case, the cutoff is no longer an intermediate mathematical construct, but instead a physical scale at which the smooth, continuous behavior of QFT breaks down.

Although it is naturally difficult to speculate about a nonexistent theory, it is generally believed that such a theory would emerge at a scale comparable to that of the Planck energy \cite{lengthscale}.
Several theories describe a spacetime made of ``quantum foam'' which violates Lorentz invariance at very high energy scales \cite{foam,foam1,foam2}, which would imply that the vacuum (which is dominated by these high-energy modes) need not be Lorentz invariant, justifying the abandonment of \Cref{item:lv}.


This abandonment of Lorentz invariance is crucial to the new approach: as discussed in \Cref{sect:ccprob}, Lorentz-invariance would require $T_{00}=-T_{ii}$ for $i=1,2,3$ (i.e.\ if energy density is positive, pressure is negative), which, from \Cref{eq:WzuOmega}, would prevent $\Omega^2$ from being positive definite and exhibiting the harmonic oscillator behavior that we describe.

\subsection{Semiclassical gravity}
\Cref{item:ev} means that it is valid to replace the right-hand side of the Einstein equation $T_{\mu\nu}$ with its expectation $\ev{T_{\mu\nu}}$.
It requires that either gravity is not in fact quantum, and the Moller-Rosenfeld approach is a complete description of reality (which is an unfavored view, see \cite{evproblems,evproblems2}), or a valid approximation in the weak-field limit (which is also not favored \cite{evproblems}).

The key development of \cite{wzu} is to replace \Cref{item:ev} with \Cref{item:ev2}, i.e.\ replace the expectation value $\ev{T_{\mu\nu}}$ with a stochastic field $T_{\mu\nu}\left( t, \vb x \right)$.
We consider this an improved approximation to a full theory of quantum gravity, as it incorporates some description of the fluctuations that we know to exist in local measurements of energy density.
If such a theory does exist, and variables like scale factor $a$ can be treated as operators, then the methods used in \cite{wzu} show that the Wigner formulation yields the approximate stochastic description used here.
Nonetheless, further work testing the rigor and applicability of these methods is required.
\change{
\subsection{Choice of Metric and the Einstein equations}
As mentioned in \Cref{sect:stochastic}, we use the simplified inhomogeneous metric (\Cref{eq:InhomMetric}) with just one degree of freedom, $a(t,\vb x)$.
This is a simplification of a more complete description, which would require a metric with ten degrees of freedom.
Our hope is that the findings presented here may extend to these more general cases, an assumption that we intend to test further in future investigations.

We also note that by using the Einstein equations at all, we continue to use unmodified general relativity. 
Thus we are also assuming that general relativity holds at all distance scales down to our cutoff scale (in fact, we assume it holds on the timescale of oscillations to $a$, i.e.\ timescales on the order of $\Lambda^2$), and that unlike some descriptions, there is no modification to Newton's constant $G$ as one ``zooms in''.
We expect that varying $G$ would result in quantitative changes to our predicted value for $H$, but the qualitative features of the model described here would persist.
}

\subsection{Number and Type of Fields}
The traditional approach as presented in \Cref{sect:ccprob} used only a single scalar field with no interactions. 
\change{Adding more fields does not change its conclusions significantly, because an increase in the number of fields (and thus uniform energy density) leads to a linear increase in cosmic acceleration, so the cosmic acceleration remains on a similar order of magnitude.
However, for this new model, the energy density and acceleration rate are no longer linearly related.}
An important goal of the current work has been to begin to relax \Cref{item:field} by testing the effects of a greater number of fields.

A single field predicts a similar outcome in the new approach as it did in the traditional approach: $H \sim 1$ with a Planck scale cutoff, and it diverges as the cutoff is increased (see \Cref{fig:morefieldsnotheory}).
However, with the new approach, adding more fields no longer worsens the problem, but instead dramatically ameliorates it!
\change{As the number of fields increases, the magnitude of the fluctuations to the energy density tends to increase more slowly than the mean increases, so that the relative magnitude of the fluctuations decreases as the inverse root of the number of fields (as per the central limit theorem).}
This causes parametric resonance to weaken, and the resultant acceleration to become smaller and smaller.

In our tests, numerical instability became more significant than the growth from $H$ below about $H\approx 10^{-6}$, i.e.\ when the acceleration $H^2$ is about $12$ orders of magnitude smaller than the traditional approach.
Because of the exponential relationship between $H$ and $\Lambda$ (which only begins past about $\sqrt{n_{\textrm{f}}}\Lambda \approx 6$), increasing the cutoff or number of fields marginally beyond this point would result in dramatically smaller acceleration, approaching the observed value $H^2=\frac{\ddot{L}}{L}\sim 10^{-120}$.

A key contribution from this work was to introduce a simple model based on the Mathieu equation, which captures many of the key features of the simulation results.
We also found that if these trends continue then we can expect $H$ to match observation when (for example) $n_{\textrm{f}}=28$ and $\Lambda=40$, or when $n_{\textrm{f}}=6000$ and $\Lambda=1$.

Of course, our description has still been restricted to massless scalar fields, and is not a complete description of the real Universe.
Our description is actually sufficient for bosonic fields, even if they are not scalar and massless.
Introducing a mass adds a term of the form $-m\phi^2$ to \Cref{eq:WzuOmega} (where $\phi$ is the field operator), which can result in $\Omega^2$ becoming negative.
But the masses of all observed particles are vastly smaller than the Planck scale, meaning this correction will have an insignicant effect on the dynamics.
Furthermore, even if a boson is not a scalar, but rather, has polarization modes like the photon, then each component still contributes to the vacuum in a manner like that of an individual scalar field.
Given the large number of bosonic field components in the standard model \footnote{$1$ from the Higgs, $2$ from the photon, $9$ from $W$ and $Z$ and $16$ from gluons, for $28$ total; see \cite{martin} Eq 401}, this amounts to a significant number of fields that our model is able to describe.

Nonetheless, this description is not sufficient for describing fermionic fields, or interactions between fields.
Fermionic fields contribute to the vacuum energy negatively, with the same magnitude (but opposite sign) as bosonic fields.
With a number of fermionic fields $ n_F$ and bosonic fields $n_B$, the mean $\ev{\Omega^2_0}$ would become $\frac{(n_B-n_F)\Lambda^4}{6\pi}$, while the variance remains related to the total number of fields (as adding more fields cannot reduce variance): $\Var(\Omega^2)=(n_B+n_F) \frac{\Lambda^8}{18 \pi^2}$.
\change{Thus the effect of adding fermionic fields is to decrease the mean and to increase the magnitude of fluctuations, increasing the strength of parametric resonance and making it harder to reach the observed $H$.
	However, so long as there is only a small probability of $\Omega^2$ fluctuating below $\Lambda$ and violating the adiabatic condition, we can still ensure weak parametric resonance rather than rapid exponential growth.
	Given large enough numbers of fields and assuming $n_B>n_F$, the chance of $\Omega^2$ fluctuating below $\Lambda$ decreases as $\exp(-k_1\frac{(n_B-n_F-k_2 \Lambda)^2}{n_B+n_F})$ for some constants $k_1$ and $k_2$.
}

With developments to our analytical description of parametric resonance, one could relate $n_F$ and $n_B$ to corresponding values of $H$, allowing a relationship between the observed $H$ and the number of fields.
Because these numbers must obviously be integers, there would be a kind of ``quantization'' of allowed $H$ values, providing both a test for this theory and a method of relating $H$ to the number of particle fields in the Universe---potentially probing dark matter fields, supersymmetric fields, etc.

\begin{acknowledgments}
	TMD acknowledges support from the ARC Centre of Excellence for All-sky Astrophysics (CAASTRO), project CE110001020.
	TCR acknowledges support from the Australian Research Council Centre of Excellence for Quantum Computation and Communication Technology (Project No. CE170100012).
	WGU thanks the Natural Science and Engineering Research Council of Canada for and the Canadian Institute for Advanced Research for funding during this research.
	SSC would also like to thank Marco Ho and Fabio Costa for useful discussions and feedback.
\end{acknowledgments}

\bibliography{refs}

\begin{thebibliography}{33}%
\makeatletter
\providecommand \@ifxundefined [1]{%
 \@ifx{#1\undefined}
}%
\providecommand \@ifnum [1]{%
 \ifnum #1\expandafter \@firstoftwo
 \else \expandafter \@secondoftwo
 \fi
}%
\providecommand \@ifx [1]{%
 \ifx #1\expandafter \@firstoftwo
 \else \expandafter \@secondoftwo
 \fi
}%
\providecommand \natexlab [1]{#1}%
\providecommand \enquote  [1]{``#1''}%
\providecommand \bibnamefont  [1]{#1}%
\providecommand \bibfnamefont [1]{#1}%
\providecommand \citenamefont [1]{#1}%
\providecommand \href@noop [0]{\@secondoftwo}%
\providecommand \href [0]{\begingroup \@sanitize@url \@href}%
\providecommand \@href[1]{\@@startlink{#1}\@@href}%
\providecommand \@@href[1]{\endgroup#1\@@endlink}%
\providecommand \@sanitize@url [0]{\catcode `\\12\catcode `\$12\catcode
  `\&12\catcode `\#12\catcode `\^12\catcode `\_12\catcode `\%12\relax}%
\providecommand \@@startlink[1]{}%
\providecommand \@@endlink[0]{}%
\providecommand \url  [0]{\begingroup\@sanitize@url \@url }%
\providecommand \@url [1]{\endgroup\@href {#1}{\urlprefix }}%
\providecommand \urlprefix  [0]{URL }%
\providecommand \Eprint [0]{\href }%
\providecommand \doibase [0]{http://dx.doi.org/}%
\providecommand \selectlanguage [0]{\@gobble}%
\providecommand \bibinfo  [0]{\@secondoftwo}%
\providecommand \bibfield  [0]{\@secondoftwo}%
\providecommand \translation [1]{[#1]}%
\providecommand \BibitemOpen [0]{}%
\providecommand \bibitemStop [0]{}%
\providecommand \bibitemNoStop [0]{.\EOS\space}%
\providecommand \EOS [0]{\spacefactor3000\relax}%
\providecommand \BibitemShut  [1]{\csname bibitem#1\endcsname}%
\let\auto@bib@innerbib\@empty
\bibitem [{\citenamefont {Wang}\ \emph {et~al.}(2017)\citenamefont {Wang},
  \citenamefont {Zhu},\ and\ \citenamefont {Unruh}}]{wzu}%
  \BibitemOpen
  \bibfield  {author} {\bibinfo {author} {\bibfnamefont {Q.}~\bibnamefont
  {Wang}}, \bibinfo {author} {\bibfnamefont {Z.}~\bibnamefont {Zhu}}, \ and\
  \bibinfo {author} {\bibfnamefont {W.~G.}\ \bibnamefont {Unruh}},\ }\href
  {\doibase 10.1103/PhysRevD.95.103504} {\bibfield  {journal} {\bibinfo
  {journal} {Phys. Rev. D}\ }\textbf {\bibinfo {volume} {95}},\ \bibinfo
  {pages} {103504} (\bibinfo {year} {2017})}\BibitemShut {NoStop}%
\bibitem [{\citenamefont {{Weinberg}}(1989)}]{weinberg}%
  \BibitemOpen
  \bibfield  {author} {\bibinfo {author} {\bibfnamefont {S.}~\bibnamefont
  {{Weinberg}}},\ }\href {\doibase 10.1103/RevModPhys.61.1} {\bibfield
  {journal} {\bibinfo  {journal} {Reviews of Modern Physics}\ }\textbf
  {\bibinfo {volume} {61}},\ \bibinfo {pages} {1} (\bibinfo {year}
  {1989})}\BibitemShut {NoStop}%
\bibitem [{\citenamefont {{Carroll}}\ \emph {et~al.}(1992)\citenamefont
  {{Carroll}}, \citenamefont {{Press}},\ and\ \citenamefont
  {{Turner}}}]{carroll}%
  \BibitemOpen
  \bibfield  {author} {\bibinfo {author} {\bibfnamefont {S.~M.}\ \bibnamefont
  {{Carroll}}}, \bibinfo {author} {\bibfnamefont {W.~H.}\ \bibnamefont
  {{Press}}}, \ and\ \bibinfo {author} {\bibfnamefont {E.~L.}\ \bibnamefont
  {{Turner}}},\ }\href {\doibase 10.1146/annurev.aa.30.090192.002435}
  {\bibfield  {journal} {\bibinfo  {journal} {ARA{\&}A}\ }\textbf {\bibinfo
  {volume} {30}},\ \bibinfo {pages} {499} (\bibinfo {year} {1992})}\BibitemShut
  {NoStop}%
\bibitem [{\citenamefont {{Martin}}(2012)}]{martin}%
  \BibitemOpen
  \bibfield  {author} {\bibinfo {author} {\bibfnamefont {J.}~\bibnamefont
  {{Martin}}},\ }\href {\doibase 10.1016/j.crhy.2012.04.008} {\bibfield
  {journal} {\bibinfo  {journal} {Comptes Rendus Physique}\ }\textbf {\bibinfo
  {volume} {13}},\ \bibinfo {pages} {566} (\bibinfo {year} {2012})}\BibitemShut
  {NoStop}%
\bibitem [{\citenamefont {Dolgov}(1997)}]{dolgov}%
  \BibitemOpen
  \bibfield  {author} {\bibinfo {author} {\bibfnamefont {A.~D.}\ \bibnamefont
  {Dolgov}},\ }in\ \href@noop {} {\emph {\bibinfo {booktitle} {{Proceedings of
  Paris}}}}\ (\bibinfo {year} {1997})\ pp.\ \bibinfo {pages} {161--175},\
  \bibinfo {note} {{Phase transitions in cosmology}},\ \Eprint
  {http://arxiv.org/abs/astro-ph/9708045} {arXiv:astro-ph/9708045 [astro-ph]}
  \BibitemShut {NoStop}%
\bibitem [{\citenamefont {Dine}(2007)}]{susskind}%
  \BibitemOpen
  \bibfield  {author} {\bibinfo {author} {\bibfnamefont {M.}~\bibnamefont
  {Dine}},\ }\href {\doibase 10.1119/1.2710490} {\bibfield  {journal} {\bibinfo
   {journal} {American Journal of Physics}\ }\textbf {\bibinfo {volume} {75}},\
  \bibinfo {pages} {382} (\bibinfo {year} {2007})}\BibitemShut {NoStop}%
\bibitem [{\citenamefont {Gr{\o}n}(2018)}]{summary}%
  \BibitemOpen
  \bibfield  {author} {\bibinfo {author} {\bibfnamefont {{\O}.~G.}\
  \bibnamefont {Gr{\o}n}},\ }\href@noop {} {\bibfield  {journal} {\bibinfo
  {journal} {Eur. J. Phys.}\ }\textbf {\bibinfo {volume} {39}},\ \bibinfo
  {pages} {043001} (\bibinfo {year} {2018})}\BibitemShut {NoStop}%
\bibitem [{\citenamefont {Rosenfeld}(1963)}]{rosenfeld}%
  \BibitemOpen
  \bibfield  {author} {\bibinfo {author} {\bibfnamefont {L.}~\bibnamefont
  {Rosenfeld}},\ }\href {\doibase https://doi.org/10.1016/0029-5582(63)90279-7}
  {\bibfield  {journal} {\bibinfo  {journal} {Nucl. Phys.}\ }\textbf {\bibinfo
  {volume} {40}},\ \bibinfo {pages} {353 } (\bibinfo {year}
  {1963})}\BibitemShut {NoStop}%
\bibitem [{\citenamefont {Parker}\ and\ \citenamefont {Toms}(2009)}]{qftcs}%
  \BibitemOpen
  \bibfield  {author} {\bibinfo {author} {\bibfnamefont {L.}~\bibnamefont
  {Parker}}\ and\ \bibinfo {author} {\bibfnamefont {D.}~\bibnamefont {Toms}},\
  }\href@noop {} {\emph {\bibinfo {title} {Quantum Field Theory in Curved
  Spacetime: Quantized Fields and Gravity}}}\ (\bibinfo  {publisher} {Cambridge
  University Press, Cambridge, England},\ \bibinfo {address} {Cambridge},\
  \bibinfo {year} {2009})\BibitemShut {NoStop}%
\bibitem [{\citenamefont {Birrell}\ and\ \citenamefont
  {Davies}(1984)}]{qftcs2}%
  \BibitemOpen
  \bibfield  {author} {\bibinfo {author} {\bibfnamefont {N.}~\bibnamefont
  {Birrell}}\ and\ \bibinfo {author} {\bibfnamefont {P.}~\bibnamefont
  {Davies}},\ }\href {https://books.google.com.au/books?id=SEnaUnrqzrUC} {\emph
  {\bibinfo {title} {Quantum Fields in Curved Space}}},\ Cambridge Monographs
  on Mathematical Physics\ (\bibinfo  {publisher} {Cambridge University Press,
  Cambridge, England},\ \bibinfo {year} {1984})\BibitemShut {NoStop}%
\bibitem [{\citenamefont {Eppley}\ and\ \citenamefont
  {Hannah}(1977)}]{evproblems2}%
  \BibitemOpen
  \bibfield  {author} {\bibinfo {author} {\bibfnamefont {K.}~\bibnamefont
  {Eppley}}\ and\ \bibinfo {author} {\bibfnamefont {E.}~\bibnamefont
  {Hannah}},\ }\href {\doibase 10.1007/BF00715241} {\bibfield  {journal}
  {\bibinfo  {journal} {Foundations of Physics}\ }\textbf {\bibinfo {volume}
  {7}},\ \bibinfo {pages} {51} (\bibinfo {year} {1977})}\BibitemShut {NoStop}%
\bibitem [{\citenamefont {Anastopoulos}\ and\ \citenamefont
  {Hu}(2014)}]{evproblems}%
  \BibitemOpen
  \bibfield  {author} {\bibinfo {author} {\bibfnamefont {C.}~\bibnamefont
  {Anastopoulos}}\ and\ \bibinfo {author} {\bibfnamefont {B.~L.}\ \bibnamefont
  {Hu}},\ }\href {http://stacks.iop.org/1367-2630/16/i=8/a=085007} {\bibfield
  {journal} {\bibinfo  {journal} {New Journal of Physics}\ }\textbf {\bibinfo
  {volume} {16}},\ \bibinfo {pages} {085007} (\bibinfo {year}
  {2014})}\BibitemShut {NoStop}%
\bibitem [{\citenamefont {Straumann}(1999)}]{straumann}%
  \BibitemOpen
  \bibfield  {author} {\bibinfo {author} {\bibfnamefont {N.}~\bibnamefont
  {Straumann}},\ }\href@noop {} {\bibfield  {journal} {\bibinfo  {journal}
  {Eur. J. Phys.}\ }\textbf {\bibinfo {volume} {20}},\ \bibinfo {pages} {419}
  (\bibinfo {year} {1999})}\BibitemShut {NoStop}%
\bibitem [{\citenamefont {Sahni}\ and\ \citenamefont
  {Starobinsky}(2000)}]{positivelambda}%
  \BibitemOpen
  \bibfield  {author} {\bibinfo {author} {\bibfnamefont {V.}~\bibnamefont
  {Sahni}}\ and\ \bibinfo {author} {\bibfnamefont {A.}~\bibnamefont
  {Starobinsky}},\ }\href {\doibase 10.1142/S0218271800000542} {\bibfield
  {journal} {\bibinfo  {journal} {International Journal of Modern Physics D}\
  }\textbf {\bibinfo {volume} {09}},\ \bibinfo {pages} {373} (\bibinfo {year}
  {2000})}\BibitemShut {NoStop}%
\bibitem [{\citenamefont {Padmanabhan}(2003)}]{padmanabhan}%
  \BibitemOpen
  \bibfield  {author} {\bibinfo {author} {\bibfnamefont {T.}~\bibnamefont
  {Padmanabhan}},\ }\href@noop {} {\bibfield  {journal} {\bibinfo  {journal}
  {Phys. Rep.}\ }\textbf {\bibinfo {volume} {380}},\ \bibinfo {pages} {235}
  (\bibinfo {year} {2003})}\BibitemShut {NoStop}%
\bibitem [{\citenamefont {Baez}\ and\ \citenamefont {Bunn}(2005)}]{attractive}%
  \BibitemOpen
  \bibfield  {author} {\bibinfo {author} {\bibfnamefont {J.~C.}\ \bibnamefont
  {Baez}}\ and\ \bibinfo {author} {\bibfnamefont {E.~F.}\ \bibnamefont
  {Bunn}},\ }\href@noop {} {\bibfield  {journal} {\bibinfo  {journal} {Am. J.
  Phys.}\ }\textbf {\bibinfo {volume} {73}},\ \bibinfo {pages} {644} (\bibinfo
  {year} {2005})}\BibitemShut {NoStop}%
\bibitem [{\citenamefont {Hu}\ and\ \citenamefont {Verdaguer}(2008)}]{hu}%
  \BibitemOpen
  \bibfield  {author} {\bibinfo {author} {\bibfnamefont {B.~L.}\ \bibnamefont
  {Hu}}\ and\ \bibinfo {author} {\bibfnamefont {E.}~\bibnamefont {Verdaguer}},\
  }\href {\doibase 10.12942/lrr-2008-3} {\bibfield  {journal} {\bibinfo
  {journal} {Living Rev. Relativity}\ }\textbf {\bibinfo {volume} {11}},\
  \bibinfo {pages} {3} (\bibinfo {year} {2008})}\BibitemShut {NoStop}%
\bibitem [{\citenamefont {{Mart{\'{\i}}n}}\ and\ \citenamefont
  {{Verdaguer}}(1999)}]{stochastic}%
  \BibitemOpen
  \bibfield  {author} {\bibinfo {author} {\bibfnamefont {R.}~\bibnamefont
  {{Mart{\'{\i}}n}}}\ and\ \bibinfo {author} {\bibfnamefont {E.}~\bibnamefont
  {{Verdaguer}}},\ }\href {\doibase 10.1103/PhysRevD.60.084008} {\bibfield
  {journal} {\bibinfo  {journal} {\prd}\ }\textbf {\bibinfo {volume} {60}},\
  \bibinfo {eid} {084008} (\bibinfo {year} {1999})}\BibitemShut {NoStop}%
\bibitem [{\citenamefont {{Akhmedov}}()}]{covariant1}%
  \BibitemOpen
  \bibfield  {author} {\bibinfo {author} {\bibfnamefont {E.~K.}\ \bibnamefont
  {{Akhmedov}}},\ }\href@noop {} {\ }\Eprint
  {http://arxiv.org/abs/arXiv:hep-th/0204048} {arXiv:hep-th/0204048}
  \BibitemShut {NoStop}%
\bibitem [{\citenamefont {{Koksma}}\ and\ \citenamefont
  {{Prokopec}}(2011)}]{covariant2}%
  \BibitemOpen
  \bibfield  {author} {\bibinfo {author} {\bibfnamefont {J.~F.}\ \bibnamefont
  {{Koksma}}}\ and\ \bibinfo {author} {\bibfnamefont {T.}~\bibnamefont
  {{Prokopec}}},\ }\href@noop {} {\  (\bibinfo {year} {2011})},\ \Eprint
  {http://arxiv.org/abs/arXiv:1105.6296 [gr-qc]} {arXiv:arXiv:1105.6296 [gr-qc]
  [gr-qc]} \BibitemShut {NoStop}%
\bibitem [{\citenamefont {{Ossola}}\ and\ \citenamefont
  {{Sirlin}}(2003)}]{covariant3}%
  \BibitemOpen
  \bibfield  {author} {\bibinfo {author} {\bibfnamefont {G.}~\bibnamefont
  {{Ossola}}}\ and\ \bibinfo {author} {\bibfnamefont {A.}~\bibnamefont
  {{Sirlin}}},\ }\href {\doibase 10.1140/epjc/s2003-01337-7} {\bibfield
  {journal} {\bibinfo  {journal} {European Physical Journal C}\ }\textbf
  {\bibinfo {volume} {31}},\ \bibinfo {pages} {165} (\bibinfo {year}
  {2003})}\BibitemShut {NoStop}%
\bibitem [{\citenamefont {Klinkhamer}(2007)}]{lengthscale}%
  \BibitemOpen
  \bibfield  {author} {\bibinfo {author} {\bibfnamefont {F.~R.}\ \bibnamefont
  {Klinkhamer}},\ }\href {\doibase 10.1134/S0021364007140019} {\bibfield
  {journal} {\bibinfo  {journal} {JETP Letters}\ }\textbf {\bibinfo {volume}
  {86}},\ \bibinfo {pages} {73} (\bibinfo {year} {2007})}\BibitemShut {NoStop}%
\bibitem [{\citenamefont {Wheeler}(1955)}]{foam}%
  \BibitemOpen
  \bibfield  {author} {\bibinfo {author} {\bibfnamefont {J.~A.}\ \bibnamefont
  {Wheeler}},\ }\href {\doibase 10.1103/PhysRev.97.511} {\bibfield  {journal}
  {\bibinfo  {journal} {Phys. Rev.}\ }\textbf {\bibinfo {volume} {97}},\
  \bibinfo {pages} {511} (\bibinfo {year} {1955})}\BibitemShut {NoStop}%
\bibitem [{\citenamefont {Crouse}(2016)}]{foam1}%
  \BibitemOpen
  \bibfield  {author} {\bibinfo {author} {\bibfnamefont {D.~T.}\ \bibnamefont
  {Crouse}},\ }\href {\doibase 10.1007/s00339-016-9853-9} {\bibfield  {journal}
  {\bibinfo  {journal} {Applied Physics A}\ }\textbf {\bibinfo {volume}
  {122}},\ \bibinfo {pages} {472} (\bibinfo {year} {2016})}\BibitemShut
  {NoStop}%
\bibitem [{\citenamefont {Caravelli}\ and\ \citenamefont
  {Markopoulou}(2012)}]{foam2}%
  \BibitemOpen
  \bibfield  {author} {\bibinfo {author} {\bibfnamefont {F.}~\bibnamefont
  {Caravelli}}\ and\ \bibinfo {author} {\bibfnamefont {F.}~\bibnamefont
  {Markopoulou}},\ }\href@noop {} {\bibfield  {journal} {\bibinfo  {journal}
  {\prd}\ }\textbf {\bibinfo {volume} {86}},\ \bibinfo {pages} {024019}
  (\bibinfo {year} {2012})}\BibitemShut {NoStop}%
\bibitem [{\citenamefont {Mazzitelli}\ and\ \citenamefont
  {Trombetta}(2017)}]{comment}%
  \BibitemOpen
  \bibfield  {author} {\bibinfo {author} {\bibfnamefont {F.~D.}\ \bibnamefont
  {Mazzitelli}}\ and\ \bibinfo {author} {\bibfnamefont {L.~G.}\ \bibnamefont
  {Trombetta}},\ }\href@noop {} {\bibfield  {journal} {\bibinfo  {journal}
  {\prd}\ }\textbf {\bibinfo {volume} {97}},\ \bibinfo {pages} {068301}
  (\bibinfo {year} {2017})}\BibitemShut {NoStop}%
\bibitem [{\citenamefont {Wang}\ and\ \citenamefont {Unruh}(2018)}]{reply}%
  \BibitemOpen
  \bibfield  {author} {\bibinfo {author} {\bibfnamefont {Q.}~\bibnamefont
  {Wang}}\ and\ \bibinfo {author} {\bibfnamefont {W.~G.}\ \bibnamefont
  {Unruh}},\ }\href {\doibase 10.1103/PhysRevD.97.068302} {\bibfield  {journal}
  {\bibinfo  {journal} {\prd}\ }\textbf {\bibinfo {volume} {97}},\ \bibinfo
  {pages} {068302} (\bibinfo {year} {2018})}\BibitemShut {NoStop}%
\bibitem [{\citenamefont {Santos}()}]{fermionic}%
  \BibitemOpen
  \bibfield  {author} {\bibinfo {author} {\bibfnamefont {E.}~\bibnamefont
  {Santos}},\ }\href@noop {} {\ }\Eprint
  {http://arxiv.org/abs/https://arxiv.org/abs/1805.03018}
  {https://arxiv.org/abs/1805.03018} \BibitemShut {NoStop}%
\bibitem [{\citenamefont {Landau}\ and\ \citenamefont
  {Lifshitz}(1976)}]{landau}%
  \BibitemOpen
  \bibfield  {author} {\bibinfo {author} {\bibfnamefont {L.~D.}\ \bibnamefont
  {Landau}}\ and\ \bibinfo {author} {\bibfnamefont {E.~M.}\ \bibnamefont
  {Lifshitz}},\ }\href@noop {} {\emph {\bibinfo {title} {Mechanics}}},\
  \bibinfo {edition} {3rd}\ ed.\ (\bibinfo  {publisher} {Butterworth-Heinemann,
  Oxford, England},\ \bibinfo {year} {1976})\BibitemShut {NoStop}%
\bibitem [{\citenamefont {Van Der~Pol}\ and\ \citenamefont
  {Strutt}(1928)}]{parresstability}%
  \BibitemOpen
  \bibfield  {author} {\bibinfo {author} {\bibfnamefont {B.}~\bibnamefont {Van
  Der~Pol}}\ and\ \bibinfo {author} {\bibfnamefont {M.}~\bibnamefont
  {Strutt}},\ }\href@noop {} {\bibfield  {journal} {\bibinfo  {journal} {The
  London, Edinburgh, and Dublin Philosophical Magazine and Journal of Science}\
  }\textbf {\bibinfo {volume} {5}},\ \bibinfo {pages} {18} (\bibinfo {year}
  {1928})}\BibitemShut {NoStop}%
\bibitem [{\citenamefont {Nijmeijer}\ and\ \citenamefont
  {Fossen}(2012)}]{parressummary}%
  \BibitemOpen
  \bibfield  {author} {\bibinfo {author} {\bibfnamefont {H.}~\bibnamefont
  {Nijmeijer}}\ and\ \bibinfo {author} {\bibfnamefont {T.~I.}\ \bibnamefont
  {Fossen}},\ }\href@noop {} {\emph {\bibinfo {title} {Parametric Resonance in
  Dynamical Systems}}}\ (\bibinfo  {publisher} {Springer, New York},\ \bibinfo
  {year} {2012})\BibitemShut {NoStop}%
\bibitem [{\citenamefont {Robnik}\ and\ \citenamefont
  {Romanovski}(2006)}]{adiabatic}%
  \BibitemOpen
  \bibfield  {author} {\bibinfo {author} {\bibfnamefont {M.}~\bibnamefont
  {Robnik}}\ and\ \bibinfo {author} {\bibfnamefont {V.}~\bibnamefont
  {Romanovski}},\ }\href@noop {} {\bibfield  {journal} {\bibinfo  {journal}
  {Open Systems and Information Dynamics}\ }\textbf {\bibinfo {volume} {13}},\
  \bibinfo {pages} {197} (\bibinfo {year} {2006})}\BibitemShut {NoStop}%
\bibitem [{\citenamefont {Alhargan}(1996)}]{mathieucomputation}%
  \BibitemOpen
  \bibfield  {author} {\bibinfo {author} {\bibfnamefont {F.~A.}\ \bibnamefont
  {Alhargan}},\ }\href@noop {} {\bibfield  {journal} {\bibinfo  {journal} {SIAM
  Rev.}\ }\textbf {\bibinfo {volume} {38}},\ \bibinfo {pages} {239} (\bibinfo
  {year} {1996})}\BibitemShut {NoStop}%
\end{thebibliography}%

\appendix
\section{Probability Distribution of $\Omega^2$} \label{sect:dist}
As explained in \Cref{sect:stochastic}, the probability distribution of $\Omega^2$ is very important in determining the validity of the adiabatic limit.
To evaluate the probability distribution, we appeal to the Wigner formulation as described in \Cref{sect:sims}.
To start with we will follow \cite{wzu}, for which the calculations are just for one field.
The Weyl transform of the nondimensionalized $\tilde \Omega^2 = \frac{3L^4}{8\pi^2} \Omega^2$ operator is given by Eq.\ (B31) of \cite{wzu}.
In the chosen nondimensionalized units used there, $\tilde{x}$ and $\tilde{p}$ are standardized normal random variables, $X(0,1)$ (we will use the notation that $X\left( \mu,s^2 \right)$ is a random variable sampled from the normal distribution with mean $\mu$ and variance $s^2$ and make use of the properties $cX(0,s^2) = X(0,c^2s^2)$ and $X(0,s^2)+X(0,s'^2) = X(0,s^2+s'^2)$ where each variable is independent). 
Then
\begin{align}
 {\tilde \Omega^2} &=  {\qty[\sum^\Lambda_{\vec n} \sqrt{n} \qty(\tilde x_{\vec n} \sin n\tilde t - \tilde p_{\vec n} \cos n \tilde t)]^2} \\
 &=  {\qty[\sum^{n_{\mathrm{max}}}_{\vec n} \sqrt{n} \qty(X(0,1) \sin n\tilde t - X(0,1) \cos n \tilde t)]^2}\\
 &=  {\qty[\sum^{n_{\mathrm{max}}}_{\vec n} \sqrt{n} \qty(X(0,\sin^2 n\tilde t)  - X(0,\cos ^2 n \tilde t) )]^2}\\
 &=  {\qty[\sum^{n_{\mathrm{max}}}_{\vec n} \sqrt{n} \qty(X(0,\sin^2 n\tilde t+ \cos ^2 n \tilde t) )]^2}  \\
 &=  {\qty[\sum^{n_{\mathrm{max}}}_{\vec n}  \qty(X(0,n) )]^2}  \\
 &=  {\qty[X\qty(0,\sum^{n_{\mathrm{max}}}_{\vec n} n) ]^2} \\
 &=  \left( \sum^{n_{\mathrm{max}}}_{\vec n}n \right) X\qty(0,1)^2 
\end{align}
Thus, generalizing to $n_{\textrm{f}}$ fields, we have
\begin{align}
	\tilde \Omega^2 &= \left( \sum^{n_{\mathrm{max}}}_{\vec n}n \right) \sum_{i=1}^{n_{\textrm{f}}} X\qty(0,1)^2 \\
	\tilde \Omega^2 &= \left( \sum^{n_{\mathrm{max}}}_{\vec n}n \right) \chi^2_{n_\textrm{f}}
	\label{<+label+>}
\end{align}
where we used the definition of $\chi^2_k$ as the sum of $k$ standard normal random variables.
To compute the sum over $\vec n$, we have (for a spherical cutoff, see next section) $\sum^{n_{\mathrm{max}}}_{\vec n} n \approx \int_0^{n_{\mathrm{max}}} n \dd\vec n =\pi n_{\mathrm{max}}^4= \pi \left( \frac{L\Lambda}{2\pi} \right)^4$.
Computing $\Omega^2$ now (noting that $L$ drops out, as it should):
\begin{align}
	\Omega^2 &=  \frac{8\pi^2}{3L^4} \tilde{\Omega}^2 \\
	\Omega^2 &=  \frac{8\pi^2}{3L^4}  \pi \left( \frac{L\Lambda}{2\pi} \right)^4 \chi^2_{n_\textrm{f}}\\
	\Omega^2 &= \frac{\Lambda^4}{6\pi}  \chi^2_{n_\textrm{f}}
	\label{<+label+>}
\end{align}
\section{Convergence Tests} \label{sect:convergence}

\begin{figure*}[ht!]
	\centering
	\includegraphics[width=\figwidth]{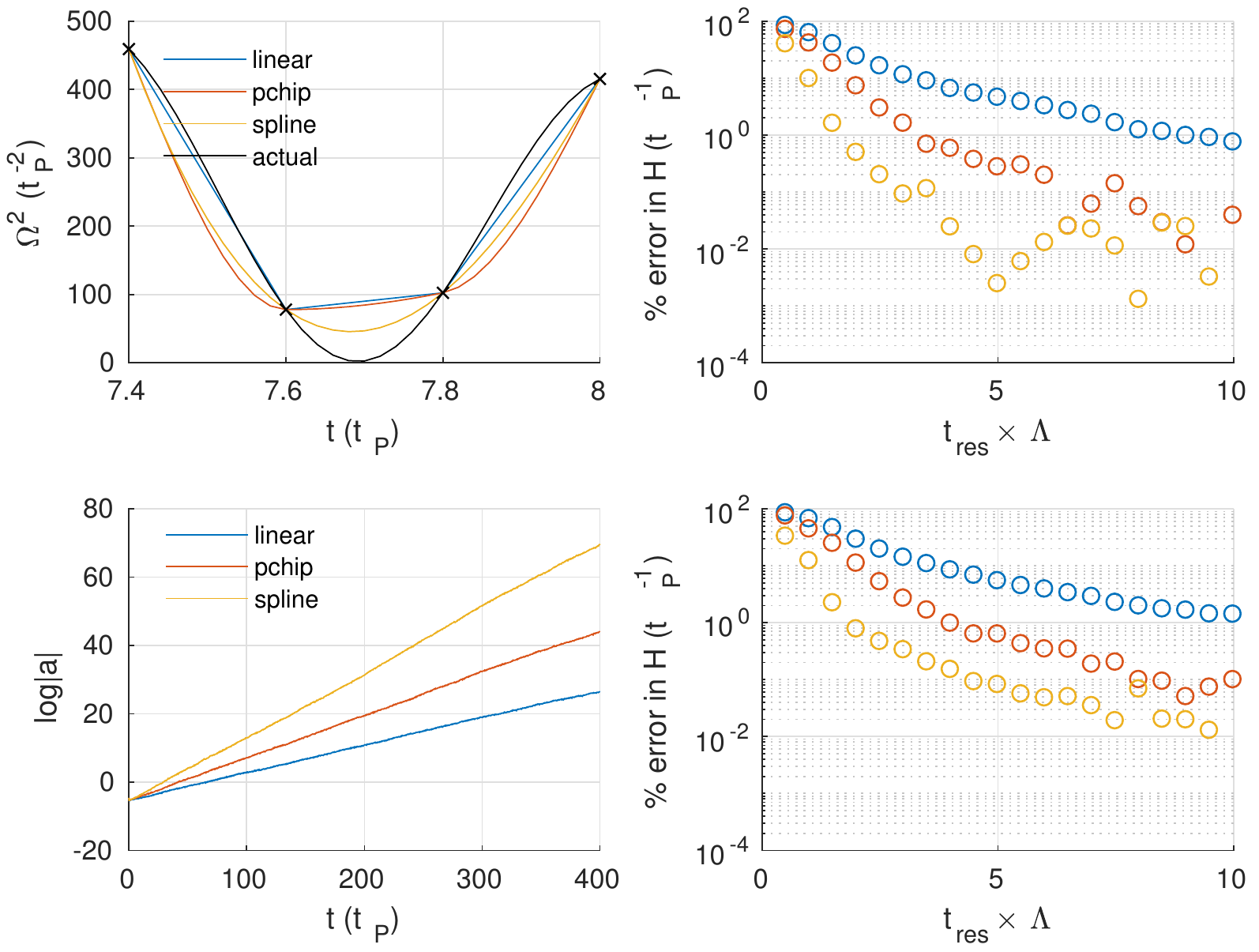}
	\caption{Here we see the effects of three different interpolation methods for $\Omega$.
	In the top left panel, the true variation in $\Omega^2$ is contrasted with three interpolation methods: linear, pchip, and spline.
	Here $\Lambda=5$, and the resolution is very coarse ($t_{\textrm{res}}=0.2$) to exaggerate the effect.
	The resultant solutions of $a(t)$ are shown in the bottom left.
	On the right, the resulting error in $H$ is shown for each interpolation method, for a range of time resolutions and two cutoff values ($\Lambda=5$ above and $\Lambda=10$ below).
	For both cutoffs, the results converge much more quickly for the spline method, indicating that $t_{\textrm{res}} = 1/3\Lambda$ is sufficient to constrain uncertainty within $1\%$.
}
	\label{fig:trestest}
\end{figure*}

\begin{figure*}[ht!]
	\centering
	\includegraphics[width=\figwidth]{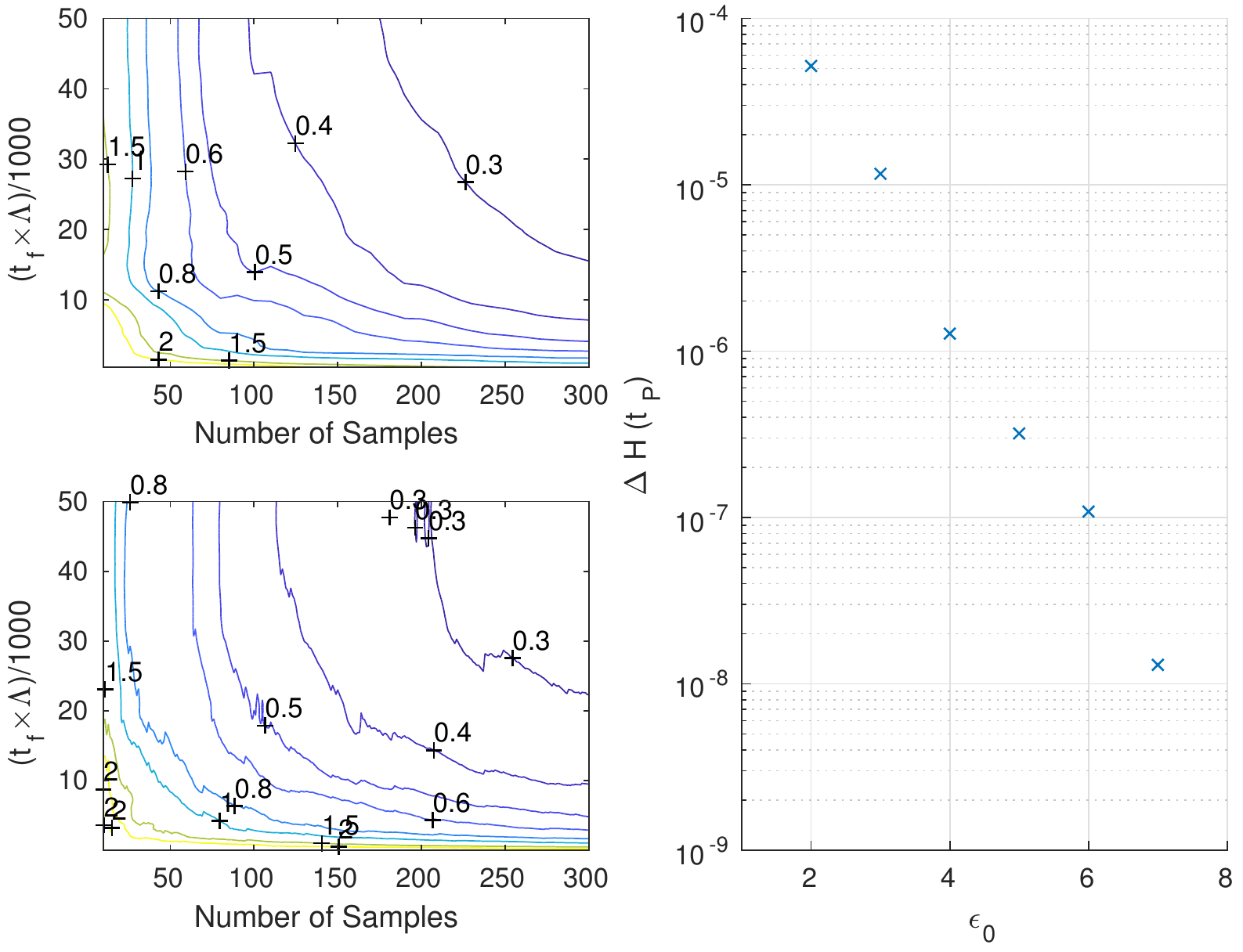}
	\caption{
		On the left, contour maps are shown for the percentage uncertainty in $H$ dependent on the number of samples ($N$) and duration of simulation ($t_f$).
		The top shows $\Lambda=5$, with $L=10$, and the bottom $\Lambda=10$ and $L=5$; this means that they will have the same maximum $|\vb n|=\frac{L\Lambda}{2\pi}$, and thus the same number of modes for consistency.
		Note that they have very similar contours, indicating that the precision depends mainly on $t_f \Lambda$ (the $y$-axis), $N$ (the $x$-axis), and $L\Lambda = 2\pi n_{\textrm{max}}$ (which is the same between the two).
		In the bottom right, we show the relationship between the absoluute uncertainty in $H$ and the ODE tolerance $\varepsilon$, parametrized by $\varepsilon_0$ as described in \Cref{eq:Epsilon}.
}
	\label{fig:tfntest}
\end{figure*}
Before beginning discussion of convergence, we must discuss our desired precision for determining $H(\Lambda)$.
Some parameters resulted in relative uncertainty, while others give absolute uncertainty values.
In order to see overall trends in $H$ with confidence, we aimed for $1\%$ uncertainty in $H$ from each parameter, or an absolute precision of $10^{-6}t_P^{-1}$, whichever was higher (here we reintroduce the unit of the Planck time $t_P$).

\subsection{Cutoff method}
Whereas a cubic cutoff was used in \cite{wzu}, i.e.\ each component $i$ satisfies $|n_i|<n_{\textrm{max}}$, we used a spherical cutoff $n=|\vb n|<n_{\mathrm{max}}$.
This difference does not affect the results greatly, except that it slightly modifies the effective $\Lambda$ being tested---with a cubic cutoff, the highest actual frequency is $\sqrt{3}\Lambda$ instead of $\Lambda$ itself.

\subsection{Interpolation method and $t_{\textrm{res}}$}
The method of interpolation turns out to be crucially important for convergence, in particular when a larger $t_{\textrm{res}}$ is being used.
We found, as shown in Figure~\ref{fig:trestest}, that of three inbuilt MATLAB interpolation methods (linear method, pchip method, and spline method), a spline interpolation converged most quickly.
It appears that the salient feature of the spline method which gives this advantage is that it extends past the upper and lower extremes of the sample points, increasing the magnitude of fluctuations of $\Omega$, as seen in the upper left panel of Figure~\ref{fig:trestest}.
The other methods underestimate the deviations to $\Omega$, which typically results in a weaker parametric resonance effect, as seen in the lower left panel of Figure~\ref{fig:trestest}.
Because oscillations of $\Omega^2$ occur on a timescale of $1/\Lambda$, as discussed in \Cref{sect:stochastic}, $t_{\textrm{res}}$ should be fixed in proportion to this time period.
From \Cref{fig:trestest}, we see that setting $t_{\textrm{res}}=1/3\Lambda$ is sufficient for uncertainty to remain within $1\%$.

\begin{figure*}[ht!]
	\centering
	\includegraphics[width=\figwidth]{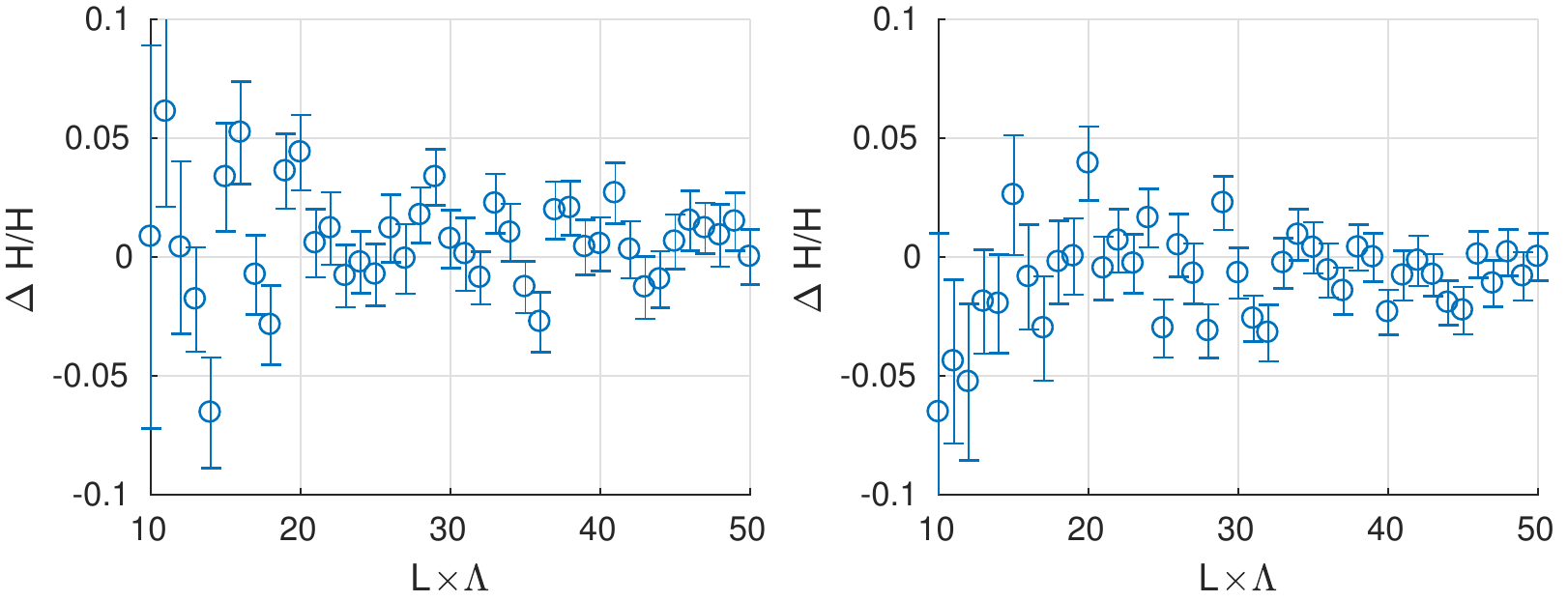}
	\caption{Here we examine the dependence of the simulations on the box length $L$, for $\Lambda=5$ on the left and $\Lambda=10$ on the right.
		Notice that once $L$ is multiplied by $\Lambda$, the convergence appears to occurs at a consistent rate between the two graphs, with $L\Lambda=50$ being sufficient for convergence within $1\%$.
}
	\label{fig:ltest}
\end{figure*}

\subsection{Dependence on ODE solver tolerance} 
The ODE solver being used, MATLAB's \begin{verb} ode45 \end{verb}, accepts a choice of relative tolerance, which we denote $\varepsilon$.
This represents the acceptable relative error in the solution per unit time, relative to its own magnitude, so it is another parameter we can tune to maximize accuracy and computational efficiency.
Within the accepted tolerance range, the amplitude of $a(t)$ may deviate from its true value (typically, it will decrease) by a fairly consistent factor each cycle, which we call $r$ (defined as a ratio, i.e.\ a perfect solution would have $r=1$). 
Thus $\log|a|$ is misestimated by an increment of $\log(r)$ per cycle, which means that as time goes on, our estimation of $\log|a|$ will linearly deviate from its true value with time.
Because $H$ is calculated as the slope of $\log|a|$, the effect of this numerical artifact will be to modify the observed $H$ by a constant $\Delta H$ compared to the correct result. 
As the number of cycles increases, i.e.\ when $\sqrt{n_{\textrm{f}}}\Lambda^2$ increases, this will occur more quickly, so we need a smaller tolerance.
For this reason, we choose the parametrization:
\begin{align}
	\varepsilon(\Lambda) = \frac{10^{-\varepsilon_0}}{\sqrt{n_{\textrm{f}}}\Lambda^2}
	\label{eq:Epsilon}
\end{align}
and investigate the dependence of $H$ on $\varepsilon_0$.
In Figure~\ref{fig:tfntest}, in the lower right, this dependence is displayed for a number of cutoffs, and we see that $\varepsilon_0 = 4$ (i.e.\ $\varepsilon=\frac{0.0001}{\sqrt{n_{\textrm{f}}}\Lambda^2}$) is enough to constrain $|\Delta H|<10^{-6} t_P^{-1}$.

\subsection{Dependence on duration of simulation and number of samples}
The duration of simulation and the number of samples are closely linked---both result in an approximate linear increase in computational difficulty (in both parts of the calculation: determining $\Omega$, in which there will be linearly more time steps or modes needed for calculation; and for determining $a(t)$ from $\Omega$, because of the number and length of differential equations needing to be solved increasing linearly).
Furthermore, both result in an inverse-square-root relationship between uncertainty in $H$ and size of $N$ or $t_f$, respectively.
This is because the total number of time steps being computed, ``$n_t$,'' is proportional to $N t_f$, and the uncertainty in estimating the average $H$ across all time steps can be computed using the usual formula, $\Delta H = \sigma/\sqrt{n_t}$, where $\sigma$ is the standard deviation.
As shown in \Cref{fig:tfntest}, $100$ samples with $t_f = 50000 \Lambda^{-1}$ is sufficient to constrain $\frac{\Delta H}{H}< 1\%$.

\subsection{Dependence on width of box}
Finally, let us consider the dependence on the width of the box $L$. 
The error for low $L$ stems from the way in which the modes are discretized in $\vb n$ space.
The sphere of allowed modes for a given field has volume given by $\frac{4\pi n_{\textrm{max}}^3}{3} = \frac{\Lambda^3 L^3}{6\pi^2}$, and because the modes are spaced as an integer lattice, the number of modes should approximate this volume.
At low $n_{\textrm{max}}$, the difference between the actual number of modes and the volume of the sphere in $\vb n$-space becomes significant, but the approximation improves for larger $n_{\textrm{max}}$.
This means that the accuracy improves for both higher $L$ and higher $\Lambda$, as is shown in Figure~\ref{fig:ltest}, and $L\Lambda=50$ is sufficient for convergence within a few percent (note that this graph also includes the error from $t_f$ and $N$, so it will not completely converge as $L\to \infty$). 

\section{The Mathieu Equation} \label{sect:mathieu}
We wish to use $\Omega^2 = \Omega_0^2\left( 1+\varepsilon \cos\gamma t \right)$, from \Cref{eq:Mathieu}, as an approximation to \Cref{eq:WzuOmega}.
There is obviously some choice about how to implement this, but we will start by ensuring that the variance and mean of the two $\Omega^2$ functions agree.
First, let us evaluate these for the Mathieu equation
\begin{align}
	\ev{\Omega^2} &=  \Omega_0^2
	\label{<+label+>}
\end{align}
\begin{align}
	\Var(\Omega^2) &=  \Omega_0^4 \left( 1 + 2\varepsilon \ev{\cos\gamma t } + \varepsilon^2 \ev{\cos^2\gamma t } \right) - \Omega_0^4 \\
	\Var(\Omega^2) &=\frac{\Omega_0^4 \varepsilon^2}{2} 
	\label{<+label+>}
\end{align}
Thus, we can determine $\varepsilon$ for our approximation by setting $\varepsilon=\frac{\sqrt{2\Var(\Omega^2)}}{\Omega_0^2}$.
As shown in \cite{wzu}, with just one field, \Cref{eq:WzuOmega} can be written in the form:
\begin{align}
	\Omega_1^2 = \Omega^2_{0,1} \left( 1 + \int_{0}^{2\Lambda} f(\gamma) \cos \gamma t + g(\gamma) \sin \gamma t\  \dd \gamma \right),
\end{align}
where $\Omega_{0,1}^2 = \frac{\Lambda^4}{6\pi}$, and $f$ and $g$ are operator-valued functions.

We can exploit the fact that the expectation values and statistical properties of $\Omega^2$ are invariant under time translations to select $t=0$ for the sake of determining variance, etc.
Then we only need $f(\gamma)$:
\begin{widetext}
\begin{align}
	f(\gamma) = -\frac{16 \pi^2}{\Lambda^4} \int_{0}^{\Lambda} \frac{\dd^3 \vb k_1 \dd^3 \vb k_2}{(2\pi)^3} \frac{\sqrt{\omega_1\omega_2}}{2} \left( a_{\vb k_1} a_{\vb k_2} + a_{\vb k_1}^{\dagger} a_{\vb k_2}^{\dagger} \right) \delta(\gamma-\omega_1-\omega_2)
\end{align}
\end{widetext}
On the vacuum, $\ev{f(\gamma)}= 0$ so $\ev{\Omega_1^2} = \Omega_{0,1}^2$, and:
\begin{align}
	\Var(\Omega_1^2) &= \Omega_{0,1}^4\left( \ev{1} + \ev{\left(  \int_{0}^{2\Lambda} f(\gamma)^2 \right) } \right) - \Omega_{0,1}^4 \\
	&= \Omega_{0,1}^4 \ev{\left(  \int_{0}^{2\Lambda} f(\gamma)^2 \right) }  
	\label{<+label+>}
\end{align}
This expectation value simplifies to exactly $2$, i.e.:
\begin{align}
	\Var(\Omega_1^2) &= 2 \Omega_{0,1}^4 
	\label{<+label+>}
\end{align}
Now, this was for one field, but because multiple fields act as multiple identical and independent variables identical to $\Omega_1^2$, we get more generally:
\begin{align}
	\Omega^2_0 = \ev{\Omega_{n_{\textrm{f}}}^2} = n_{\textrm{f}} \Omega_{0,1}^2 = \frac{n_{\textrm{f}}\Lambda^4}{6\pi}
	\label{<+label+>}
\end{align}
\begin{align}
	\Var(\Omega_{n_{\textrm{f}}}^2) &= n_{\textrm{f}} \Var(\Omega^2_1) \\
	&= 2 n_{\textrm{f}} \Omega_{0,1}^4 \\
	\varepsilon &=  \frac{\sqrt{2\left( 2 n_{\textrm{f}} \Omega_{0,1}^4 \right)}}{n_{\textrm{f}} \Omega_{0,1}^2} = \frac{2}{\sqrt{n_{\textrm{f}}}}
	\label{<+label+>}
\end{align}
With these values set, then, we have $r = \frac{2\Omega_0}{\gamma} = \frac{2\sqrt{n_{\textrm{f}}}\Lambda^2}{\gamma \sqrt{6\pi}}$, for a variety of $\gamma$ values between $0$ and $2\Lambda$ as per \Cref{fig:PowerSpectrum}.
At the highest $\gamma$, this corresponds to $r = \Lambda \sqrt{\frac{n_{\textrm{f}}}{6\pi}}$. 
To make it into the form of \Cref{eq:Mathieu}, we should choose the ``most important'' $\gamma^*$ and then replace $f(\gamma)$ with a Dirac delta function $\delta(\gamma-\gamma^*)$.
It is important to account for two factors: the strength of the resonance (as we do not want to select a $\gamma$ with no resonance at all, i.e.,\ a white region of \Cref{fig:stability}), and also the amplitude of $\Omega^2$'s oscillations at that frequency, as given by (see \Cref{fig:PowerSpectrum}):
\begin{align}
P(\gamma) &= \ev{f(\gamma)^2} \\
&=
\begin{cases}
	\frac{2}{35 \Lambda}  \left( \frac{\gamma}{\Lambda} \right)^7, & 0\leq \gamma \leq \Lambda \\
\frac{2}{35 \Lambda} \left(- \frac{\gamma^7}{\Lambda^7} +70  \frac{\gamma^3}{\Lambda^3} -168  \frac{\gamma^2}{\Lambda^2} +140  \frac{\gamma}{\Lambda} -40\right) , & \Lambda \leq \gamma \leq 2\Lambda \\
\end{cases}
	\label{<+label+>}
\end{align}
We can quantify the resonance using the Mathieu exponent $H$, which is computed according to \cite{mathieucomputation} using:
\begin{align}
	H(\gamma) = H\left(r= \frac{2\Lambda^2}{\gamma}\sqrt{\frac{n_{\textrm{f}}}{6\pi}},\varepsilon= \frac{2}{\sqrt{n_{\textrm{f}}}} \right)
	\label{<+label+>}
\end{align}
We then choose $\gamma^*$ such that it maximizes the product $F(\gamma)=P(\gamma)H(\gamma)$.

Now, we should consider the dimensions of these quantities in order to normalize $F(\gamma^*)$ and quantify the actual growth of the $H$ in the simulations.
Because of the way that the Mathieu functions are computed, $H(\gamma)$ quantifies the growth in nondimensionalized units of time, specifically, $e^H$ is the growth factor per time unit $2/\gamma$.
Given that our actual $H$ is a frequency, to rescale it appropriately we need to multiply by $\gamma^*/2$ to reinstate units of frequency.

The units of $P(\gamma^*)$ are inverse frequency, because it is integrated to give a normalized total power.
Thus we should multiply by the width of frequencies which all contribute to excite the resonance---i.e.\ multiply by the width of the relevant resonance band from \Cref{fig:stability}.
For example, if $\gamma_{\textrm{min}}$ and $\gamma_{\textrm{max}}$ denote the lowest and highest $\gamma$ which lie in the resonance band, then we multiply by $\Delta \gamma = \gamma_{\textrm{max}} - \gamma_{\textrm{min}}$.
All in all, we have
\begin{align}
	H_{\textrm{estimate}} = P(\gamma^*)  H\left(\gamma^* \right)\frac{\gamma^* }{2} \Delta \gamma 
\end{align}
This is the estimate used in \Cref{fig:fixedcutoff}.
\end{document}